\PassOptionsToPackage{unicode}{hyperref}
\PassOptionsToPackage{hyphens}{url}
\PassOptionsToPackage{dvipsnames,svgnames,x11names}{xcolor}
\documentclass[
  12pt,
  letterpaper,
  DIV=11,
  numbers=noendperiod]{scrartcl}
\usepackage{xcolor}
\usepackage{amsmath,amssymb}
\usepackage{iftex}
\ifPDFTeX
  \usepackage[T1]{fontenc}
  \usepackage[utf8]{inputenc}
  \usepackage{textcomp} % provide euro and other symbols
\else % if luatex or xetex
  \usepackage{unicode-math} % this also loads fontspec
  \defaultfontfeatures{Scale=MatchLowercase}
  \defaultfontfeatures[\rmfamily]{Ligatures=TeX,Scale=1}
\fi
\usepackage{lmodern}
\ifPDFTeX\else
\fi
\IfFileExists{upquote.sty}{\usepackage{upquote}}{}
\IfFileExists{microtype.sty}{% use microtype if available
  \usepackage[]{microtype}
  \UseMicrotypeSet[protrusion]{basicmath} % disable protrusion for tt fonts
}{}
\makeatletter
\@ifundefined{KOMAClassName}{% if non-KOMA class
  \IfFileExists{parskip.sty}{%
    \usepackage{parskip}
  }{% else
    \setlength{\parindent}{0pt}
    \setlength{\parskip}{6pt plus 2pt minus 1pt}}
}{% if KOMA class
  \KOMAoptions{parskip=half}}
\makeatother
\makeatletter
\ifx\paragraph\undefined\else
  \let\oldparagraph\paragraph
  \renewcommand{\paragraph}{
    \@ifstar
      \xxxParagraphStar
      \xxxParagraphNoStar
  }
  \newcommand{\xxxParagraphStar}[1]{\oldparagraph*{#1}\mbox{}}
  \newcommand{\xxxParagraphNoStar}[1]{\oldparagraph{#1}\mbox{}}
\fi
\ifx\subparagraph\undefined\else
  \let\oldsubparagraph\subparagraph
  \renewcommand{\subparagraph}{
    \@ifstar
      \xxxSubParagraphStar
      \xxxSubParagraphNoStar
  }
  \newcommand{\xxxSubParagraphStar}[1]{\oldsubparagraph*{#1}\mbox{}}
  \newcommand{\xxxSubParagraphNoStar}[1]{\oldsubparagraph{#1}\mbox{}}
\fi
\makeatother

\usepackage{color}
\usepackage{fancyvrb}

\DefineVerbatimEnvironment{Highlighting}{Verbatim}{commandchars=\\\{\}}
\usepackage{framed}
\definecolor{shadecolor}{RGB}{241,243,245}
\newenvironment{Shaded}{\begin{snugshade}}{\end{snugshade}}

\newcommand{\FunctionTok}[1]{\textcolor[rgb]{0.28,0.35,0.67}{#1}}

\newcommand{\NormalTok}[1]{\textcolor[rgb]{0.00,0.23,0.31}{#1}}

\newcommand{\OtherTok}[1]{\textcolor[rgb]{0.00,0.23,0.31}{#1}}

\usepackage{longtable,booktabs,array}
\usepackage{calc} % for calculating minipage widths
\usepackage{etoolbox}
\makeatletter
\patchcmd\longtable{\par}{\if@noskipsec\mbox{}\fi\par}{}{}
\makeatother
\IfFileExists{footnotehyper.sty}{\usepackage{footnotehyper}}{\usepackage{footnote}}
\makesavenoteenv{longtable}
\usepackage{graphicx}
\makeatletter
\newsavebox\pandoc@box
\newcommand*\pandocbounded[1]{% scales image to fit in text height/width
  \sbox\pandoc@box{#1}%
  \Gscale@div\@tempa{\textheight}{\dimexpr\ht\pandoc@box+\dp\pandoc@box\relax}%
  \Gscale@div\@tempb{\linewidth}{\wd\pandoc@box}%
  \ifdim\@tempb\p@<\@tempa\p@\let\@tempa\@tempb\fi% select the smaller of both
  \ifdim\@tempa\p@<\p@\scalebox{\@tempa}{\usebox\pandoc@box}%
  \else\usebox{\pandoc@box}%
  \fi%
}
\def\fps@figure{htbp}
\makeatother

\NewDocumentCommand\citeproctext{}{}
\NewDocumentCommand\citeproc{mm}{%
  \begingroup\def\citeproctext{#2}\cite{#1}\endgroup}
\makeatletter
 \let\@cite@ofmt\@firstofone
 \def\@biblabel#1{}
 \def\@cite#1#2{{#1\if@tempswa , #2\fi}}
\makeatother
\newlength{\cslhangindent}
\newlength{\csllabelwidth}
\newenvironment{CSLReferences}[2] % #1 hanging-indent, #2 entry-spacing
 {\begin{list}{}{%
  \setlength{\itemindent}{0pt}
  \setlength{\leftmargin}{0pt}
  \setlength{\parsep}{0pt}
  \ifodd #1
   \setlength{\leftmargin}{\cslhangindent}
   \setlength{\itemindent}{-1\cslhangindent}
  \fi
  \setlength{\itemsep}{#2\baselineskip}}}
 {\end{list}}
\usepackage{calc}

\providecommand{\tightlist}{%
  \setlength{\itemsep}{0pt}\setlength{\parskip}{0pt}}

\usepackage{tcolorbox}
\usepackage{amssymb}
\usepackage{yfonts}
\usepackage{bm}

\newtcolorbox{greybox}{
  colback=white,
  colframe=blue,
  coltext=black,
  boxsep=5pt,
  arc=4pt}

\KOMAoption{captions}{tableheading}
\makeatletter
\@ifpackageloaded{caption}{}{\usepackage{caption}}
\AtBeginDocument{%
\ifdefined\contentsname
  \renewcommand*\contentsname{Table of contents}
\else
  \newcommand\contentsname{Table of contents}
\fi
\ifdefined\listfigurename
  \renewcommand*\listfigurename{List of Figures}
\else
  \newcommand\listfigurename{List of Figures}
\fi
\ifdefined\listtablename
  \renewcommand*\listtablename{List of Tables}
\else
  \newcommand\listtablename{List of Tables}
\fi
\ifdefined\figurename
  \renewcommand*\figurename{Figure}
\else
  \newcommand\figurename{Figure}
\fi
\ifdefined\tablename
  \renewcommand*\tablename{Table}
\else
  \newcommand\tablename{Table}
\fi
}
\@ifpackageloaded{float}{}{\usepackage{float}}
\floatstyle{ruled}
\@ifundefined{c@chapter}{\newfloat{codelisting}{h}{lop}}{\newfloat{codelisting}{h}{lop}[chapter]}
\floatname{codelisting}{Listing}

\makeatother
\makeatletter
\@ifpackageloaded{caption}{}{\usepackage{caption}}
\@ifpackageloaded{subcaption}{}{\usepackage{subcaption}}
\makeatother
\usepackage{bookmark}
\IfFileExists{xurl.sty}{\usepackage{xurl}}{} % add URL line breaks if available
\hypersetup{
  pdftitle={Yet Another Smacof - Square Symmetric Case},
  pdfauthor={Jan de Leeuw},
  colorlinks=true,
  linkcolor={blue},
  filecolor={Maroon},
  citecolor={Blue},
  urlcolor={Blue},
  pdfcreator={LaTeX via pandoc}}

\title{Yet Another Smacof - Square Symmetric Case}
\author{Jan de Leeuw}
\date{November 28, 2025}
\begin{document}
\maketitle
\begin{abstract}
We rewrite the metric/nonmetric and weighted/unweighted versions of the
smacof program for square symmetric data as one monolithic C program. R
is used for taking care of the data and parameter setup, the I/O, and of
issuing a single call to \emph{.C()} to start the computations. This
makes this new \emph{smacofSS()} program five to fifty times as fast
(for our examples) as the \emph{smacofSym()} function from the R smacof
package. Utilities for various initial configurations and plots are
included in the package. Examples are included to compare output and
time for the R and C versions and to illustrate the various plots.
\end{abstract}

\renewcommand*\contentsname{Table of contents}
{
\hypersetup{linkcolor=}
\setcounter{tocdepth}{3}
\tableofcontents
}

\textbf{Note:} This is a working manuscript which may be
expanded/updated frequently. All suggestions for improvement are
welcome. All Rmd, tex, html, pdf, R, and C files are in the public
domain. Attribution will be appreciated, but is not required. All files
can be found at \url{https://github.com/deleeuw/smacofFlat}

\section{Introduction}\label{introduction}

In Multidimensional Scaling (MDS) we start with \(n\) \emph{objects}.
Objects can be anything: people, animals, molecules, locations, time
points, stimuli, political parties, and so on. We have information about
the \emph{similarities} or \emph{dissimilarities} of some or all of the
\(\binom{n}{2}\) pairs of objects. Say we have information about \(m\)
dissimilarities \(\delta_k\), with \(k=1,\cdots m\). Thus each index
\(k\) refers to a pair of indices \((i,j)\), with \(1\leq i\leq n\) and
\(1\leq j\leq n\). The information is of the form \(\delta\in\Delta\),
where \(\Delta\) is a known subset of the non-negative orthant of
\(\mathbb{R}^m\). If the dissimilarities are known numbers and
\(\Delta\) has only a single element the MDS problem is \emph{metric},
in all other cases the problem is \emph{non-metric}.

In MDS we want to map the objects into \emph{points} in
\(p\)-dimensional Euclidean space in such a way that the distances
\(d_k\) between the points approximate the dissimilarities \(\delta_k\).
The quality of the approximation is given by the least squares loss
function\footnote{The symbol $:=$ is used for definitions.}
\begin{equation}
\sigma(X,\Delta):=\frac{\sum_{k=1}^mw_k(\delta_k-d_k(X))^2}{\sum_{k=1}^mw_k\delta_k^2},
\label{eq-stressdef}
\end{equation} Traditionally, this loss function is called \emph{stress}
(Kruskal (\citeproc{ref-kruskal_64a}{1964a}), Kruskal
(\citeproc{ref-kruskal_64b}{1964b})). It is true that in Kruskal's paper
there are no weights and the sum of squares of the distances is in the
denominator. It is shown in Kruskal and Carroll
(\citeproc{ref-kruskal_carroll_69}{1969}), and in more detail in De
Leeuw (\citeproc{ref-deleeuw_U_75a}{1975}), that the solution of the MDS
problem for the different denominators are the same, up a scale constant
which is irrelevant for MDS.

In definition \eqref{eq-stressdef}

\begin{itemize}
\tightlist
\item
  \(\delta_k\) are \(m\) non-negative \emph{pseudo-distances};
\item
  \(w_k\) are \(m\) positive \emph{weights};
\item
  \(X\) is an \(n\times p\) \emph{configuration}, with coordinates of
  \(n\) \emph{points} in \(\mathbb{R}^p\);
\item
  \(d_k(X)\) are the (Euclidean) \emph{distances} between the rows of
  the configuration matrix.
\end{itemize}

The (unconstrained, least squares, square, symmetric, Euclidean,
\(p-\)dimensional) MDS problem is to minimize \(\sigma\) of
\eqref{eq-stressdef} over all \(n\times p\) configurations and over the
set \(\Delta\in\mathbb{R}^m_+\) of non-negative pseudo-distances.

\section{MDS Data Structure}\label{mds-data-structure}

Data management and algorithm initialization is handled by R (R Core
Team (\citeproc{ref-r_core_team_25}{2025})). We start with an R object
of class \emph{dist} containing dissimilarities. Here is a small example
of order four.

\begin{verbatim}
  1 2 3
2 1    
3 3 1  
4 2 3 1
\end{verbatim}

Turn this into MDS data with the utility \emph{makeMDSData()}, which
creates an object of class \emph{smacofSSData}.

\begin{Shaded}
\begin{Highlighting}[]
\NormalTok{smallData }\OtherTok{\textless{}{-}} \FunctionTok{makeMDSData}\NormalTok{(small)}
\FunctionTok{print}\NormalTok{(smallData)}
\end{Highlighting}
\end{Shaded}

\begin{verbatim}
$iind
[1] 2 3 4 4 3 4

$jind
[1] 1 2 3 1 1 2

$delta
[1] 1 1 1 2 3 3

$blocks
[1] 3 0 0 1 2 0

$weights
[1] 1 1 1 1 1 1

$nobj
[1] 4

$ndat
[1] 6

attr(,"class")
[1] "smacofSSData"
\end{verbatim}

Note that the data in column \emph{delta} are increasing, and that
\emph{blocks} are \emph{tie-blocks}, i.e.~they indicate how many
elements are equal to the first element of the block. Also, the
\emph{weights} are always there in some form or another, even for MDS
analyses that are \emph{unweighted} (i.e.~when all \(w_k\) are equal).
An object of class \emph{smacofSSData} is \emph{complete} if all
\(\binom{n}{2}\) dissimilarities are present. \emph{makeMDSData()} can
handle missing data and nontrivial weights. If our example is

\begin{verbatim}
   1  2  3
2 NA      
3  3  1   
4 NA  3  1
\end{verbatim}

and we add weights, also of class \emph{dist},

\begin{verbatim}
  1 2 3
2 1    
3 1 3  
4 2 1 0
\end{verbatim}

then our \emph{smacofSSData} object becomes

\begin{Shaded}
\begin{Highlighting}[]
\NormalTok{smallData }\OtherTok{\textless{}{-}} \FunctionTok{makeMDSData}\NormalTok{(smallMissing, smallWeights)}
\FunctionTok{print}\NormalTok{(smallData)}
\end{Highlighting}
\end{Shaded}

\begin{verbatim}
$iind
[1] 3 3 4

$jind
[1] 2 1 2

$delta
[1] 1 3 3

$blocks
[1] 1 2 0

$weights
[1] 3 1 1

$nobj
[1] 4

$ndat
[1] 3

attr(,"class")
[1] "smacofSSData"
\end{verbatim}

In the dist objects \emph{delta} and \emph{weights} the missing data are
coded as zero weights, or as dissimilarities and/or weights that are
\emph{NA}. Zero dissimilarities do not indicate missing data. Thus
weights are always strictly positive and missing data do not enter into
the data object at all.

These conventions make it possible to also handle rectangular
off-diagonal data, such as this \emph{delta} and \emph{weights}.

\begin{verbatim}
  1 2 3 4 5 6
2 0          
3 0 0        
4 0 0 0      
5 1 3 1 1    
6 2 1 3 3 0  
7 3 1 2 3 0 0
\end{verbatim}

\begin{verbatim}
  1 2 3 4 5 6
2 0          
3 0 0        
4 0 0 0      
5 1 1 1 1    
6 1 1 1 1 0  
7 1 1 1 1 0 0
\end{verbatim}

Now \emph{makeMDSData()} gives

\begin{verbatim}
$iind
 [1] 5 6 7 5 5 6 7 7 5 6 6 7

$jind
 [1] 1 2 2 3 4 1 3 1 2 3 4 4

$delta
 [1] 1 1 1 1 1 2 2 3 3 3 3 3

$blocks
 [1] 5 0 0 0 0 2 0 5 0 0 0 0

$weights
 [1] 1 1 1 1 1 1 1 1 1 1 1 1

$nobj
[1] 7

$ndat
[1] 12

attr(,"class")
[1] "smacofSSData"
\end{verbatim}

Note however that handling rectangular data with square symmetric MDS is
inefficient, and it is better to use smacof programs specifically
designed for rectangular data.

It is of course also possible to construct \emph{smacofSSData} objects
in other ways, and to edit the objects generated by
\emph{makeMDSData()}, for instance by deleting/adding observations or
transforming weights/dissimilarities. As long as the conventions are
obeyed that no index pair \((i,j)\) occurs more than once, that the
dissimilarities remain sorted, and that the tie-blocks faithfully
reflect ties in the sorted dissimilarities. We do need \(i\neq j\), but
it is not necessary that always \(i>j\).

\section{Computation}\label{computation}

We minimize stress from \eqref{eq-stressdef} by minimizing its numerator
\(\smash{\sum_{k=1}^m w_k(\delta_k-d_k(X))^2}\) over the \(n\times p\)
configurations \(X\) and over \(\delta\in\Delta\cap\mathcal{S}\), where
\(\mathcal{S}\) is set of all vectors in \(\mathbb{R}^m\) with
\(\smash{\sum_{k=1}^mw_k\delta_k^2=1}\). In De Leeuw
(\citeproc{ref-deleeuw_U_75a}{1975}) this is called \emph{explicit
normalization} (of the disparities) to contrast it with the
\emph{implicit normalization} in \eqref{eq-stressdef}. It is also shown
in De Leeuw (\citeproc{ref-deleeuw_U_75a}{1975}) that explicit and
implicit normalization give the same solution for \(\delta\) and \(X\)
up to a scale constant. See also Bauschke, Bui, and Wang
(\citeproc{ref-bauschke_bui_wang_18}{2018}).

The default initial configuration for the iterations is the classical
scaling solution, with missing data imputed as average non-missing
dissimilarities. Smacof algorithms for MDS use \emph{Alternating Least
Squares (ALS)} iterations to minimize stress. ALS is a form of block
relaxation applied to least squares loss functions. Specific cases of
ALS have been around for a long time. But as a general class of
techniques it was introduced in De Leeuw
(\citeproc{ref-deleeuw_R_68d}{1968}), where it was also given its name.
If there are two blocks of variables we alternate finding the optimum
for the variables in the first block, with those in the second block
fixed at their current values, and finding the optimum for the variables
in the second block, with those in the first block fixed at their new
current values. In the terminology of Guttman
(\citeproc{ref-guttman_68}{1968}) this strategy defines a
\emph{two-phase algorithm}. ALS is especially relevant for non-metric
scaling, in which the two sets of unknowns \(X\) and \(\delta\) are
nicely separated.

Finding the optimal \(X\) for given \(\delta\) means solving a metric
MDS problem. The vector \(\delta\) is fixed throughout the computations,
in other words \(\Delta\) is a singleton, a set with one element. To
solve the optimum \(X\) subproblem we use \emph{majorization}, which was
introduced to MDS by De Leeuw (\citeproc{ref-deleeuw_C_77}{1977}). The
majorization details are now in many places, for example in Borg and
Groenen (\citeproc{ref-borg_groenen_05}{2005}) or Groenen and Van de
Velden (\citeproc{ref-groenen_vandevelden_16}{2016}). Although
majorization in metric MDS does not use ALS, it is a form of block
relaxation (De Leeuw (\citeproc{ref-deleeuw_C_94c}{1994})) that also
proceeds by solving a sequence of relatively simple least squares
problems.

In the non-metric case finding the optimum \(\delta\) for fixed \(X\),
and thus fixed \(d(X)\), is a \emph{monotone regression} problem, i.e.~a
linear least squares problem with monotonicity constraints on the
parameters. There is a huge literature on monotone regression and the
\emph{Pool Adjacent Violators Algorithm (PAVA)} reviewed in De Leeuw,
Hornik, and Mair (\citeproc{ref-deleeuw_hornik_mair_A_09}{2009}). In the
MDS context it suffices to refer to Kruskal
(\citeproc{ref-kruskal_64b}{1964b}).

In our implementation of smacof we alternate one majorization step with
one monotone regression step. Thus in the first phase we make only one
step towards the conditional minimum over \(X\), while in the second
phase we go all the way to the conditional minimum over \(\delta\) .
There may be some improvement possible by making more majorization steps
in the first phase, or by using the over-relaxation step proposed by De
Leeuw and Heiser (\citeproc{ref-deleeuw_heiser_C_80}{1980}). However,
these possible accelerations have not been implemented yet.

In the non-metrix case our algorithm is quite different from that in
Kruskal (\citeproc{ref-kruskal_64a}{1964a}), Kruskal
(\citeproc{ref-kruskal_64b}{1964b}). Kruskal defines his non-metric
stress as \begin{equation}
\sigma(X):=\min_{\delta\in\Delta}\frac{\sum_{k=1}^mw_k(\delta_k-d_k(X))^2}{\sum_{k=1}^mw_kd_k^2(X)}.
\end{equation} This stress is a function of \(X\) only, which is then
minimized by a gradient method with a complicated step-size procedure.
In Guttman's terminology this is a \emph{one-phase} procedure. Even in
the metric case Kruskal's steps differs from smacof's majorization
steps, which are gradient steps with a constant step-size that
guarantees monotone convergence to a local minimum from any starting
point.

In the code used in this paper \emph{smacofSS()} is an R function in the
R front end \emph{smacofSS.R} that reads the parameters and data of the
problem and then loads a shared library, called
\emph{smacofSSEngine.so}, which contains the compiled code of the C
routine \emph{smacofSSEngine()}. The C routine \emph{smacofSSEngine()}
alternates calls to the compiled C routines \emph{smacofSSMajorize()}
and \emph{smacofSSMonotone()}. Because of this modularity the same code
can be used with small modifications for alternative MDS methods such as
McGee's Elastic Scaling and Sammon Mapping (De Leeuw
(\citeproc{ref-deleeuw_E_25f}{2025a}), De Leeuw
(\citeproc{ref-deleeuw_E_25g}{2025b})).

The github repository has the R and C code, some utilities in R, and a
small Makefile for the shared library.

\section{Arguments}\label{arguments}

The smacofSS() function has the following arguments, with default
values,

\begin{enumerate}
\def\labelenumi{\arabic{enumi}.}
\tightlist
\item
  \emph{theData}, a \emph{smacofSSData} object.
\item
  \emph{ndim=2}, dimensionality of MDS analysis.
\item
  \emph{xinit=NULL}, initial configuration, NULL or an \emph{nobj} by
  \emph{ndim} matrix.
\item
  \emph{ties=1}, ties approach, 1, 2, or 3 for primary, secondary,
  tertiary.
\item
  \emph{weighted = FALSE}, TRUE/FALSE for weighted/unweighted least
  squares.
\item
  \emph{ordinal = FALSE}, FALSE for numerical,TRUE for ordinal.
\item
  \emph{itmax = 1000}, maximum number of iterations.
\item
  \emph{eps = 1e-10}, if stress changes less than \emph{eps}, stop.
\item
  \emph{digits = 10}, digits stress print if verbose is TRUE,
\item
  \emph{width = 12}, width stress print if verbose is TRUE,
\item
  \emph{verbose = FALSE}, TRUE/FALSE print stress for each iteration to
  \emph{stdout}.
\end{enumerate}

\section{Value}\label{value}

The list of objects returned by \emph{smacofSS()} mimics, as much as
possible, the list returned by the function \emph{smacofSym()} from the
smacof package.

\begin{enumerate}
\def\labelenumi{\arabic{enumi}.}
\tightlist
\item
  \emph{delta}, dissimilarities, vector of length \(m\).
\item
  \emph{dhat}, final pseudo-distances, vector of length \(m\).
\item
  \emph{confdist}, final distances, vector of length \(m\).
\item
  \emph{conf}, final configuration, \(n\times p\) matrix.
\item
  \emph{weightmat}, weights, vector of length \(m\).
\item
  \emph{stress}, final stress.
\item
  \emph{ndim}, number of dimensions.
\item
  \emph{init}, initial configuration, \(n\times p\) matrix..
\item
  \emph{niter}, number of iterations.
\item
  \emph{nobj}, number of objects.
\item
  \emph{iind}, row indices, vector of length \(m\).
\item
  \emph{jind}, column indices, vector of length \(m\).
\item
  \emph{weighted}, was the analysis weighted.
\item
  \emph{ordinal}, was the analysis ordinal (and if so, which tie
  approach).
\end{enumerate}

One important special case should be kept in mind. In the ordinal case
with the primary approach to ties, the data are (potentially) reordered
within tie blocks in each iteration. In each iteration we have to
reorder the indices \emph{iind}, \emph{jind}, and in the weighted case
the weights. There is no need to reorder delta and the blocks, because
the only ordering changes are within blocks. Ultimately this means that
if smacofSS() returns a list, then in that list \emph{iind},
\emph{jind}, and in the weighted case \emph{weightmat} will be ordered
differently from the corresponding columns in the MDS data structure. No
reordering is going on using the secondary and tertiary approaches.

\section{Details}\label{sec-details}

\subsection{Interface C and R}\label{sec-interface}

The C routine \emph{smacofSSEngine()} by itself is a complete smacof MDS
programs. It would be easy to write a main program in C that do the same
job as the R frontend, and compile it to a stand-alone executable.

The C code uses the \emph{.C()} calling conventions and is otherwise not
dependent on R in any way. It is important to note that the \emph{.C()}
call from R is only executed one time in a \emph{smacofSS} job, when
loading the shared library, and after that there is only compiled code
running until the end of the job. In other words, everything done in R
is either front-end or back-end, and is only done once. Earlier versions
of the R/C program had R also managing the iterations. In each iteration
there was a \emph{.C()} call to update the configuration and in the
ordinal case an additional \emph{.C()} call to update the
pseudo-distances. This turned out to be unsatisfactory, because it did
not give enough speedup compared to the \emph{smacofSym()} in the
\emph{smacof} package on CRAN (De Leeuw and Mair
(\citeproc{ref-deleeuw_mair_A_09c}{2009}), Mair, Groenen, and De Leeuw
(\citeproc{ref-mair_groenen_deleeuw_A_22}{2022})).

\subsection{Majorization}\label{majorization}

The majorization updates are of the form \(X\leftarrow V^+B(X)X\), where
\(V\) and \(B(X)\) are symmetric doubly-centered matrices. Matrix \(V\)
is constant and has off-diagonal elements \(-w_{ij}\), matrix \(B(X)\)
has off-diagonal elements \(-w_{ij}\delta_{ij}/d_{ij}(X)\). Matrix
\(V^+\) is a generalized inverse of \(V\). For details see, for example,
De Leeuw and Heiser (\citeproc{ref-deleeuw_heiser_C_80}{1980}) or De
Leeuw (\citeproc{ref-deleeuw_A_88b}{1988}). If the weights are
irreducible, which we can assume without loss of generality (De Leeuw
(\citeproc{ref-deleeuw_C_77}{1977})), then \begin{equation}
V^+=(V+\frac{1}{n}ee')^{-1}-\frac{1}{n}ee'
\end{equation} can be used to efficiently compute the Moore-Penrose
inverse. In the unweighted case we can simply set
\(\smash{V^+=\frac{1}{n}I}\).

If \emph{weighted} is true then smacof requires two multiplications of
dense symmetric matrices of order \(n\) in each iteration. This is the
most expensive part of the calculations. We handle these multiplications
in the C routine \emph{smacofSSMajorize()}, which is called by
\emph{smacofSSEngine()}, by storing and computing only the part below
the diagonal and by using double-centering and symmetry in the
calculations. The Moore-Penrose inverse of the weighting matrix \(V\) is
also needed (only once) and it is computed by the C routine
\emph{smacofMPInverseV()} using only the lower diagonal elements of
\(V\) with symmetric sweeping.

\subsection{Monotone Regression}\label{sec-monotone}

If \emph{ordinal} is TRUE the monotone regression in each iteration uses
the C version of the \emph{monotone()} algorithm of Busing
(\citeproc{ref-busing_22}{2022}). The \emph{monotone()} algorithm is
called in three separate C routines \emph{primaryApproach()},
\emph{secondaryApproach()}, and \emph{tertiaryApproach()}, for which the
code is in \emph{smacofIsotone.c}. One of the three tie approaches is
called in each iteration by the \emph{smacofSSMonotone()} routine (which
in turn is called by \emph{smacofSSEngine()}).

\subsection{Initial Configuration}\label{sec-initial}

The initial configuration routines are written in R. They are

\begin{enumerate}
\def\labelenumi{\arabic{enumi}.}
\tightlist
\item
  \emph{smacofTorgerson(theData, ndim)}. Returns initial configuration
  in \(nobj\times ndim\) matrix using classical MDS.
\item
  \emph{smacofGuttman(theData, ndim)}. Returns the Guttman-Lingoes
  initial configuration in \(nobj\times ndim\) matrix.
\item
  \emph{smacofElegant(theData, ndim)}. Returns the ``elegant'' initial
  configuration in \(nobj\times ndim\) matrix.
\end{enumerate}

They all need the first \emph{ndim} eigenvalues with the corresponding
eigenvalues. We use \emph{eigs-sym()} from the RSpectra package (Qiu and
Mei (\citeproc{ref-qiu_mei_24}{2024})) for this. We can also generate a
random initial configuration with \emph{smacofRandomConfiguration()} in
\emph{smacofAuxiliaries.R}. Random configurations are not really useful
for data analysis, but can be used in theoretical studies of convergence
and local minima.

Each of the three routines is actually a fairly complete metric MDS
program that takes a \emph{smacofSSData} object as its argument and
returns a value similar in structure to the \emph{smacofSSResult} object
returned by \emph{smacofSS()}. Thus the returned object can be used
directly the plotting routines of Section~\ref{sec-plotting}.

In terms of strategy we suggest that in an MDS analysis the researcher
first computes one or more initial configurations (which are all
independent MDS solutions anyway, except for the random configuration),
and then give one or more of these initial configurations as the
\emph{xinit} argument to \emph{smacofSS()}.

The initial configuration routines all are based on finding a solution
with a good value of \emph{sstress}, defined in the same way as stress,
but with squared distances and dissimilarities. Thus sstress is
\begin{equation}
\sigma(X,\Delta):=\frac{\sum_{k=1}^mw_k(\delta_k^2-d_k^2(X))^2}{\sum_{k=1}^mw_k\delta_k^4}.
\label{eq-sstressdef}
\end{equation} The \emph{smacofSSResult} returned by the initial
configuration routines has an sstress value, instead of a stress value.
Of source the stress value of the sulution will be revealed when it is
used as an initial configuration in \emph{smacofSS()}. Also, when
\emph{smacofSSResult} from an sstress routine is used for a Shepardplot,
squared distances and squared dhats are plotted against unsquared deltas
so the plot has a quadratic shape.

\subsubsection{Torgerson}\label{sec-torgerson}

The default initial configuration is set to NULL for \emph{smacofSS()}.
This is mean to emphasize the importance of choosing an initial
configuration. But to make life easier for the user if \emph{xinit} is
null, then an initial configuration is computed by the
\emph{smacofTorgerson()} routine in \emph{smacofAuxiliaries.R}.
\emph{smacofTorgerson()} is classical MDS (Torgerson
(\citeproc{ref-torgerson_58}{1958})), with the missing dissimilarities
imputed using the average non-missing dissimilarity.

The Torgerson initial configuration works best for complete and
unweighted data, but may also be satisfactory for a small number of
randomly missing data and for weights that do not vary a great deal.
Also remember that in the case of a really bad fit classical MDS may run
into the problem of negative eigenvalues. This will give the initial
configuration a dimension less than \(p\), and since smacof iterations
never increase dimensionality this will mean the final configuration
will also have dimension less than \(p\). Thus the Torgerson initial
configuration is unsuitable, for example, for full-dimensional scaling
(De Leeuw, Groenen, and Mair
(\citeproc{ref-deleeuw_groenen_mair_E_16e}{2016})).

\subsubsection{Guttman}\label{sec-guttman}

The Guttman-Lingoes initial configuration (Guttman
(\citeproc{ref-guttman_68}{1968})), which handles missing data and
unequal weights, is provided as \emph{smacofGuttman()}. It is closely
related to Hayashi's Quantification Method IV (Takane
(\citeproc{ref-takane_77}{1977})) and unlike the Torgerson method does
not have a negative eigenvalue problem.

Write sstress as \begin{equation}
\sigma(X)=K-2\sum_{k=1}^mw_k\delta_k^2d_k^2(X)+\sum_{k=1}^mw_kd_k^4(X).
\end{equation} By homogeneity this is scale-equivalent to maximizing the
middle term over all configurations for which the third term is one. The
middle term is equal to \(\text{tr}\ X'BX\), where \(B\) is
doubly-centered with off-diagonal elements \(-d_{ij}^2(X)\). Note that
\(B\) is positive semi-definite. The third term
\(\smash{\sum_{k=1}^mw_kd_k^4(X)}\) is a homogeneous quartic in \(X\),
or equivalently a homogeneous quadratic in the elements of \(C=XX'\). In
the \emph{elegant} algorithm, discussed in Section~\ref{sec-elegant}, we
majorize this quadratic to arrive at a simpler unweighted quadratic in
\(C\). In the Guttman initial configuration we maximize the middle term
using the alternative quartic constraint
\(\text{tr}\ (X'X)^2=\text{tr}\ C^2=1\). Note that the \(B\) takes
weights and missing data into account, but the normalization constraint
does not. The stationary equations are \(BX=X(X'X)\) and thus the
maximum is attained for \(\smash{X=K\Lambda^\frac12}\), where
\(\Lambda\) are the \(p\) dominant eigenvalues of \(B\) and \(K\) are
the corresponding eigenvalues.

\subsubsection{Elegant}\label{sec-elegant}

Our most expensive initial configuration routine is
\emph{smacofElegant()}, which implements an optimized version of De
Leeuw's 1975 ``elegant'' algorithm, as described recently in De Leeuw
(\citeproc{ref-deleeuw_E_25c}{2025c}). \emph{smacofElegant()} is a
stand-alone iterative metric scaling method, implemented in R. Using
\emph{smacofElegant()} means a lot of computation just to obtain an
initial configuration for \emph{smacofSS()}. On the other hand it is
crucial in MDS to have an initial configuration which is as good as
possible, because that is the best guarantee against local minima.

\subsubsection{Full-dimensional Scaling}\label{full-dimensional-scaling}

Finally, we have the option to uses the \(p\) dominant dimensions of the
\emph{full-dimensiponal} scaling solution as initial configuration. We
know that full-dimensional scaling converges to the global minimum of
stress in \(n-1\) dimensions, and actually to the global minimum for all
\(p\geq r\), where \(r\) is the \emph{Gower rank} of the data (De Leeuw
(\citeproc{ref-deleeuw_E_16k}{2016})). This will provide a good initial
estimate if \(r\) is close to \(p\).

\section{Utilities}\label{utilities}

\subsection{Data Utilities}\label{data-utilities}

There are four functions in smacofDataUtilities.R

\begin{enumerate}
\def\labelenumi{\arabic{enumi}.}
\tightlist
\item
  \emph{makeMDSData(delta, weights = NULL)}. Returns \emph{smacofSSData}
  object from dist objects.
\item
  \emph{fromMDSData(theData)}. Returns \emph{dist} objects of
  dissimilarities and weights from \emph{smacofSSData} object.
\item
  \emph{matrixPrint(x, digits = 6, width = 8, format = ``f'', flag =
  ``+'')}. Formats and prints a matrix.
\item
  \emph{smacofRandomConfiguration(theData, ndim = 2)}. Random initial
  configuration.
\end{enumerate}

\subsection{Plotting}\label{sec-plotting}

There are three plot routines in the file \emph{smacofPlots.R}.

\begin{enumerate}
\def\labelenumi{\arabic{enumi}.}
\tightlist
\item
  \emph{smacofShepardPlot(h, main = ``ShepardPlot'', fitlines = TRUE,
  colline = ``RED'', colpoint = ``BLUE'', resolution = 100, lwd = 2, cex
  = 1, pch = 16)}
\item
  \emph{smacofConfigurationPlot(h, main = ``ConfigurationPlot'', labels
  = NULL, dim1 = 1, dim2 = 2, pch = 16, col = ``RED'', cex = 1)}
\item
  \emph{smacofDistDhatPlot(h, fitlines = TRUE, colline = ``RED'',
  colpoint = ``BLUE'', main = ``Dist-Dhat Plot'', cex = 1, lwd = 2, pch
  = 16)}
\end{enumerate}

The Shepard plot has the original dissimilarities on the horizontal
axis, and both the pseudo-distances and the distances on the vertical
axis. Pseudo-distances are plotted as points of color \emph{colline},
and are connected by a line of color \emph{colline}. Distances are
plotted as points if color \emph{colpoint}. If \emph{fitlines} is TRUE
then black vertical lines from the distance point to the pseudo-distance
point are drawn.

The configuration plot plots two dimensions \emph{dim1} and \emph{dim2}
of the configuration. If \emph{labels} is NULL points are drawn with
symbol \emph{pch}, otherwise a vector of labels is used.

A DistDhat-plot has distances on the horizontal axis and
pseudo-distances on the vertical axes. Points are drawn using symbol
\emph{pch} in color \emph{colpoint}. The line through the origin with
slope one is drawn in color \emph{colline}. If \emph{fitlines} is true
then black lines from the points to their orthogonal projections on the
line are drawn.

\section{Example Data Sets}\label{example-data-sets}

There are a number of example data sets in the directory smacofSSData.
They are mostly of the 20th century boomer type: small and collected
from the aggregated judgments of human subjects. All data are available
both as dist objects and as MDS data structures. The Wish and Iris data
are not used in our computations but are included as examples of MDS
data structures.

\subsection{Ekman Data}\label{ekman-data}

Similarity data between 14 colors from Ekman
(\citeproc{ref-ekman_54}{1954}). Rating scale similarity judgments
averaged over subjects, linearly converted to unit interval
dissimilarities.

\subsection{Morse Data}\label{morse-data}

Dissimilarity between the Morse codes of 36 letters and numbers from
Rothkopf (\citeproc{ref-rothkopf_57}{1957}). Confusion probabilities
transformed to dissimilarities.

\subsection{Gruijter Data}\label{gruijter-data}

Dissimilarities of nine Dutch political parties in 1967, from De
Gruijter (\citeproc{ref-degruijter_67}{1967}). Collected using the
method of triads and averaged over 100 subjects.

\subsection{Wish Data}\label{wish-data}

Rating scale similarities between 12 nations, collected by Wish
(\citeproc{ref-wish_71}{1971}). Averaged over subjects, subtracted from
maximum scale value to form dissimilarities.

\subsection{Iris Data}\label{iris-data}

Classical iris data from Anderson (\citeproc{ref-anderson_36}{1936}) via
Fisher (\citeproc{ref-fisher_36}{1936}). Euclidean distances over four
measurements on 150 irises.

\section{Comparisons}\label{comparisons}

\subsection{Initial}\label{initial}

In this section we use the Ekman and Gruijter data to compare the three
different initial configurations. Parameters are all set at defaults,
which means a two-dimensional solution with unweighted and numerical
options.

We start with the Ekman data, which have a very good fit in two
dimensions. From the Torgerson initial configuration we need 25
iterations to converge to stress 0.0172132. From the Guttman initial
configuration stress is 0.0172132 after 27 iterations. And from elegant
we use 23 iterations to arrive at 0.0172132.

We repeat this for the Gruijter data, which have a poor fit in two
dimensions. From the Torgerson initial configuration we need 318
iterations to converge to stress 0.0446034. From the Guttman initial
configuration stress is 0.0444297 after 315 iterations. And from elegant
we use 322 iterations to arrive at 0.0444297.

For both Ekman and Gruijter it seems that smacofSS() arrives at the same
solution, no matter which initial configuration we use. There is some
small difference in the number of iterations that are required, and it
seems the Torgerson initial configuration holds its own.
\emph{smacofElegant()} is expensive and does not seem to deliver any
extra efficiency. So, on the basis of the limited experience reported
here, we do not recommend it for general use.

\subsection{Outcome}\label{outcome}

In this section we compare the output (final stress, number of
iterations) of smacofSym() and smacofSS() using the Ekman and Morse
data. Note that smacofSym() reports the square root of the final stress,
following Kruskal (\citeproc{ref-kruskal_64a}{1964a}), so for comparison
purposes we square it again. All runs are started with the
two-dimensional Torgerson solution and have the stop criteria \emph{eps}
equal to \ensuremath{10^{-10}} and \emph{itmax} equal to 1000 (except
when using the tertiary approach when \emph{itmax} is 10000). In the
weighted case the weights for Ekman are \(w_k=\delta_k^2\), for Morse
they are \(w_k=\delta_k^{-1}\).

\begin{longtable}[]{@{}lcccc@{}}
\caption{Comparison smacofSym() and smacofSS() results Ekman
data}\tabularnewline
\toprule\noalign{}
& Sym stress & Sym niter & SS stress & SS niter \\
\midrule\noalign{}
\endfirsthead
\toprule\noalign{}
& Sym stress & Sym niter & SS stress & SS niter \\
\midrule\noalign{}
\endhead
\bottomrule\noalign{}
\endlastfoot
unweighted numerical & 0.0172132 & 25 & 0.0172132 & 25 \\
unweighted ordinal ties = 1 & 0.0005337 & 103 & 0.0005337 & 103 \\
unweighted ordinal ties = 2 & 0.0009977 & 51 & 0.0009977 & 51 \\
unweighted ordinal ties = 3 & 0.0000001 & 2556 & 0.0000001 & 2556 \\
weighted numerical & 0.0105187 & 22 & 0.0086555 & 21 \\
weighted ordinal ties = 1 & 0.0003205 & 78 & 0.0003205 & 78 \\
weighted ordinal ties = 2 & 0.0007063 & 64 & 0.0007063 & 64 \\
weighted ordinal ties = 3 & 0.0000002 & 4650 & 0.0000002 & 4650 \\
\end{longtable}

\begin{longtable}[]{@{}lcccc@{}}
\caption{Comparison smacofSym() and smacofSS() results Morse
data}\tabularnewline
\toprule\noalign{}
& Sym stress & Sym niter & SS stress & SS niter \\
\midrule\noalign{}
\endfirsthead
\toprule\noalign{}
& Sym stress & Sym niter & SS stress & SS niter \\
\midrule\noalign{}
\endhead
\bottomrule\noalign{}
\endlastfoot
unweighted numerical & 0.0899492 & 238 & 0.0899492 & 238 \\
unweighted ordinal ties = 1 & 0.0326557 & 143 & 0.0326557 & 143 \\
unweighted ordinal ties = 2 & 0.0406405 & 135 & 0.0406405 & 135 \\
unweighted ordinal ties = 3 & 0.0000018 & 351 & 0.0000018 & 351 \\
weighted numerical & 0.0977124 & 317 & 0.1152147 & 320 \\
weighted ordinal ties = 1 & 0.0346208 & 117 & 0.0346208 & 117 \\
weighted ordinal ties = 2 & 0.0425777 & 99 & 0.0425777 & 99 \\
weighted ordinal ties = 3 & 0.0000025 & 289 & 0.0000025 & 289 \\
\end{longtable}

The conclusion is clear. For both the Ekman and the Morse example the
results of \emph{smacofSym()} and \emph{smacofSS()} are identical. Not
only the final stress (and thus presumably the final configuration) but
also the number of iterations (and thus presumably all iterations). This
boosts our confidence in the correctness of both programs.

\subsection{Time}\label{time}

The \emph{microbenchmark} package (Mersmann
(\citeproc{ref-mersmann_24}{2024})) is mainly intended to time small
pieces of code, not complete programs. Nevertheless we will use it to
compare \emph{smacofSym()} and \emph{smacofSS()}, until something more
appropriate comes along. We again use the Ekman and Morse data with
default options. From the microbenchmark output we find the median time
over 100 runs each of the two programs.

\begin{longtable}[]{@{}lccc@{}}
\caption{Comparison smacofSym() and smacofSS() running times Ekman
Data}\tabularnewline
\toprule\noalign{}
& SS time & Sym time & ratio Sym/SS \\
\midrule\noalign{}
\endfirsthead
\toprule\noalign{}
& SS time & Sym time & ratio Sym/SS \\
\midrule\noalign{}
\endhead
\bottomrule\noalign{}
\endlastfoot
unweighted numerical & 84337.0 & 987423.5 & 11.70807 \\
unweighted ordinal ties = 1 & 99999.0 & 5530551.5 & 55.30607 \\
unweighted ordinal ties = 2 & 85198.0 & 1918492.5 & 22.51805 \\
unweighted ordinal ties = 3 & 168305.0 & 101810421.0 & 604.91620 \\
weighted numerical & 87207.0 & 917375.0 & 10.51951 \\
weighted ordinal ties = 1 & 105349.5 & 4332859.5 & 41.12843 \\
weighted ordinal ties = 2 & 89995.0 & 2398131.0 & 26.64738 \\
weighted ordinal ties = 3 & 173040.5 & 188350843.0 & 1088.47838 \\
\end{longtable}

The entries in the table for the tertiary approach are suspect, because
tertiary uses a huge number of iterations to arrive at a perfect
solution. In some of the microbenchmark runs it may actually use the
full 10000 iterations. From the better conditioned numerical and ordinal
solutions we see that \emph{smacofSS()} is between 10 and 55 times as
fast as \emph{smacofSym()}.

\begin{longtable}[]{@{}lccc@{}}
\caption{Comparison smacofSym() and smacofSS() running times Morse
data}\tabularnewline
\toprule\noalign{}
& SS time & Sym time & ratio Sym/SS \\
\midrule\noalign{}
\endfirsthead
\toprule\noalign{}
& SS time & Sym time & ratio Sym/SS \\
\midrule\noalign{}
\endhead
\bottomrule\noalign{}
\endlastfoot
unweighted numerical & 1095950 & 15995289 & 14.594901 \\
unweighted ordinal ties = 1 & 4059308 & 15567884 & 3.835109 \\
unweighted ordinal ties = 2 & 1360646 & 9056510 & 6.656035 \\
unweighted ordinal ties = 3 & 3152634 & 25514054 & 8.092934 \\
weighted numerical & 1393877 & 21178530 & 15.193973 \\
weighted ordinal ties = 1 & 3405398 & 11178630 & 3.282620 \\
weighted ordinal ties = 2 & 1110157 & 6335607 & 5.706947 \\
weighted ordinal ties = 3 & 2593230 & 20426159 & 7.876726 \\
\end{longtable}

We see that for the Morse data \emph{smacofSS()} is 3 to 15 times faster
than \emph{smacofSym()}. The difference is largest in the numerical
case, and it is also larger for unweighted than for weighted. Strangely
enough for the ordinal options the weighted case is faster than the
corresponding non-weighted case, although a majorization iteration in
the weighted case is more expensive. This is undoubtedly because the
outcome results in the previous section show that fewer iterations are
used in the weighted case. We also see that secondary is faster than
primary and tertiary (which is rather well-behaved in the Morse example)
is faster than secondary.

We see that (in these examples) the ratio of execution times for
\emph{smacofSym()} and \emph{smacofSS()} is larger for the smaller
example. We verify this for the Iris example, which has 11175
dissimilarities. For unweighted/numerical the Sym/SS ratio is 2.5, for
unweighted/ordinal it is 1.06. Execution times are influenced by many
factors. In the majorization step \emph{smacofSym()} uses full matrices,
which is wasteful in terms of storage, but allows access to the
efficient matrix multiplication routines in R (which are already in
compiled C code). \emph{smacofSS()} has the more flexible
\emph{smacofSSData} structure, which requires less storage but more
index manipulation. Both \emph{smacofSym()} and \emph{smacofSS()} use
efficient monotone regression routines in C, so the more time the
programs spend in the monotone regression step the more equal execution
times will be. A reasonable hypothesis is that using \emph{smacofSSData}
will be relatively advantageous if there are many missing data and if
there are weights. But more precise profiling will be a worthwhile
future project.

The two examples are very different because Ekman has an excellent fit,
while Morse has a much poorer one. In both examples the tertiary
approach to ties leads to a perfect fit in Ekman and a near perfect fit
in Morse. Morse (630 dissimilarities) is quite a bit bigger than Ekman
(91 dissimilarities) and Morse has more ties (68 tie-blocks of average
size about 9) than Ekman (47 tie-blocks of average size about 2). Except
for tertiary Morse requires considerably more iterations than Ekman.

Note, by the way, that we have used microbenchmark in such a way that
smacofSS() is always run after the corresponding smacofSym(), and that
the eight different basic combinations or weighted and ordinal are
always run in the same order.

\section{Plots}\label{plots}

To show some plots we repeat eight possible analyses (numeric/ordinal,
where ordinal has three tie approaches, and weighted/unweighted) using
the Ekman and Morse data. The number of iterations and the final stress
have already been reported in the comparisons section. Both data sets
have a fairly large number of ties, so we expect the choice of the ties
approach to have some effect.

The next eight pages have Shepard plots and DistDhat plots for each of
the four unweighted analyses. No plots are given for the weighted
analyses, because they would illustrate essentially the same points. We
made the plots large so they show some detail. Shepard plots have
\emph{fitlines} equal to TRUE, DistDhat plots have \emph{fitlines} equal
to TRUE for the Ekman data and FALSE for the Morse data. The sum of the
squared lengths of the vertical fitlines in the Shepard plots is the
stress. In the DistDhat plots the sum of squares of distances of the
points to their orthogonal projections on the line is also the stress.
Thus we can see from the plots where the largest residuals are, although
the plot does not show which pair of points the fitlines correspond
with. That information can easily be obtained from the numerical output.

The Ekman example has an exceptionally good fit, even in the numerical
case. Still, allowing for ordinal transformations gives a major
improvement. Especially the DistDhat plots in the ordinal case show the
different ways of handling tie blocks. The tertiary approach gives what
is essentially a perfect fit, but the Shepard plot shows that in order
to achieve this deviations from monotonicity are required.

The Morse example has a bad numerical fit, and the improvements by the
ordinal options are huge. There are no fitlines in the DistDhat plots,
because that would mainly result in big black blobs. As in the Ekman
example the tertiary approach gives close to perfect fit, at the cost of
many deviations from monotonicity.

\begin{center}
\pandocbounded{\includegraphics[keepaspectratio]{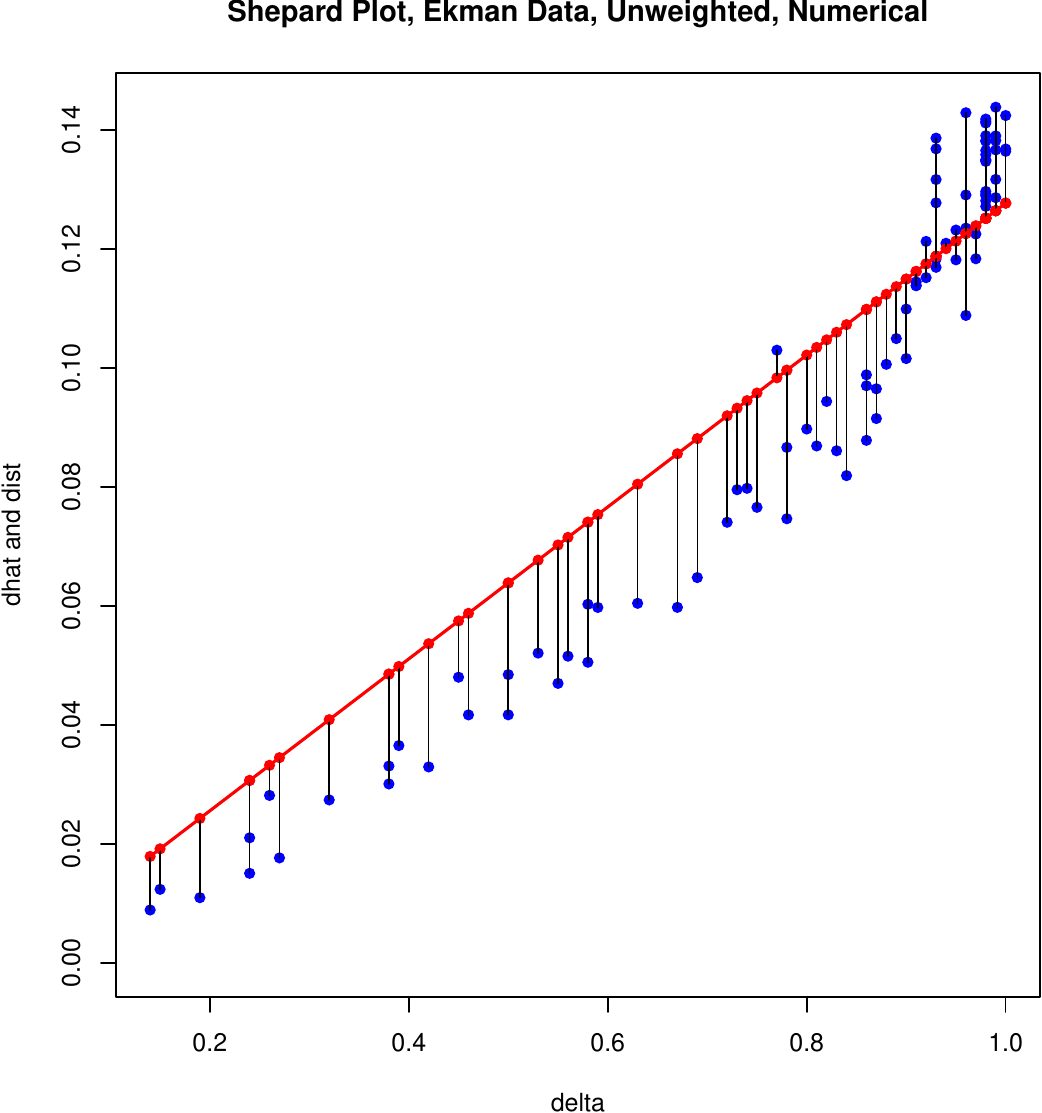}}
\end{center}

\begin{center}
\pandocbounded{\includegraphics[keepaspectratio]{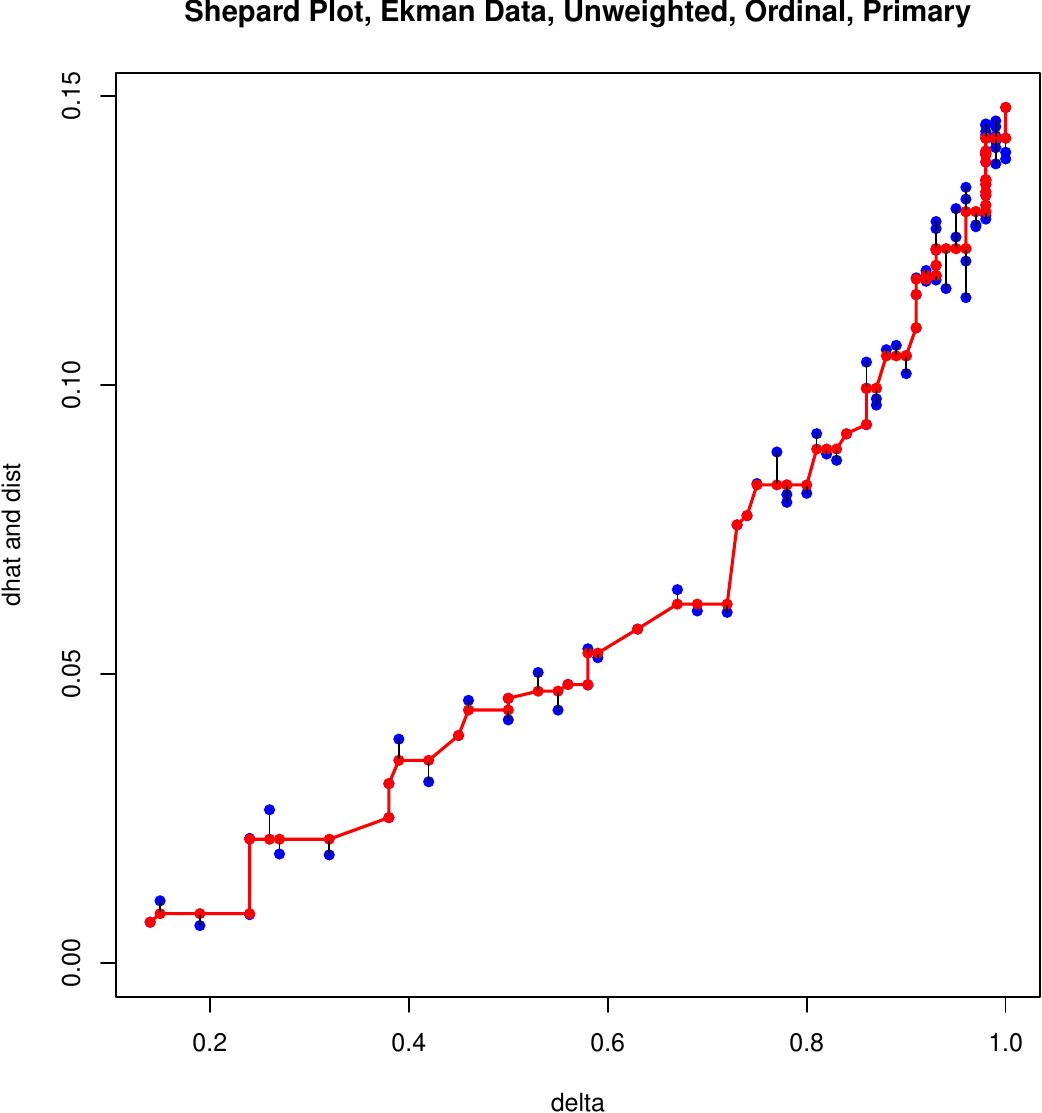}}
\end{center}

\begin{center}
\pandocbounded{\includegraphics[keepaspectratio]{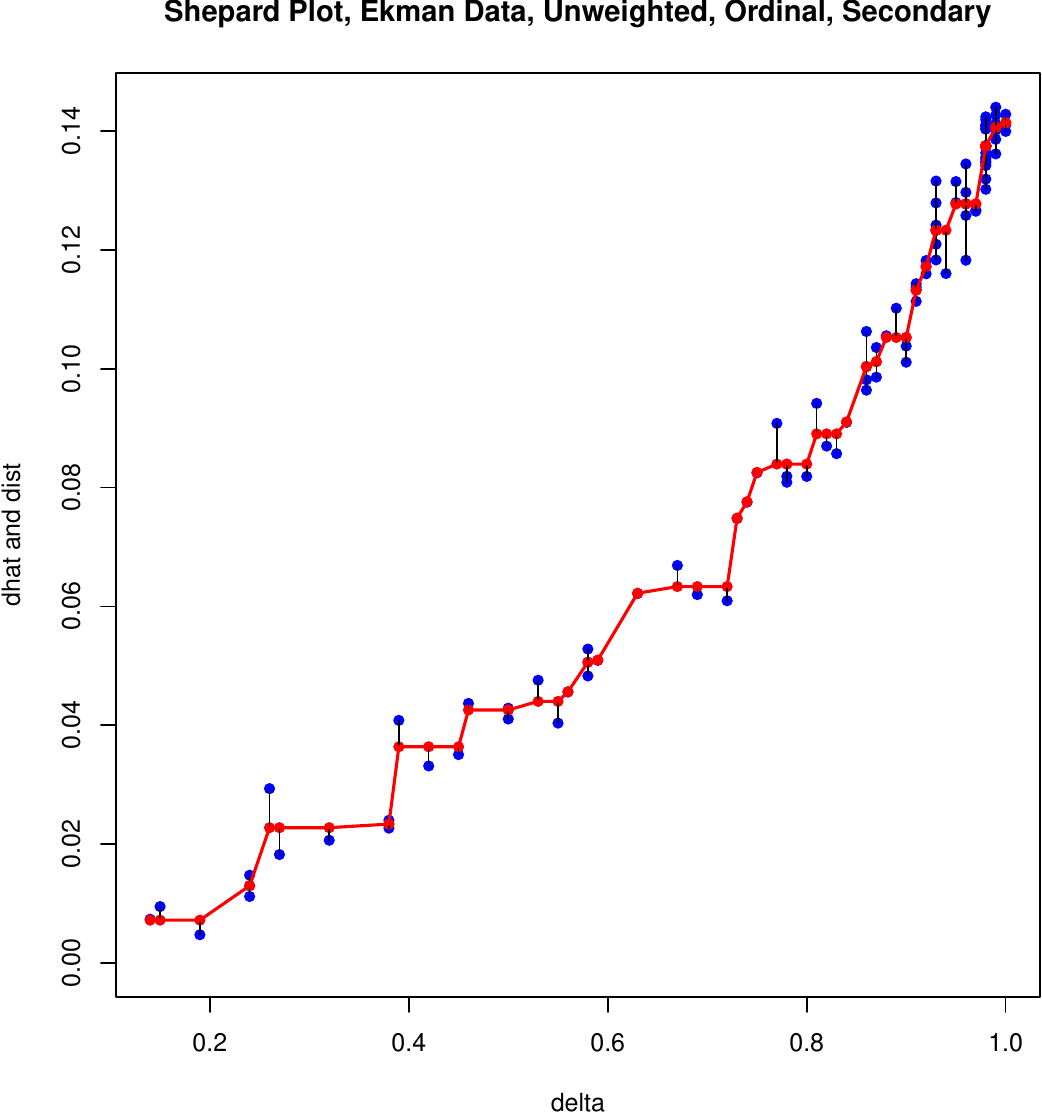}}
\end{center}

\begin{center}
\pandocbounded{\includegraphics[keepaspectratio]{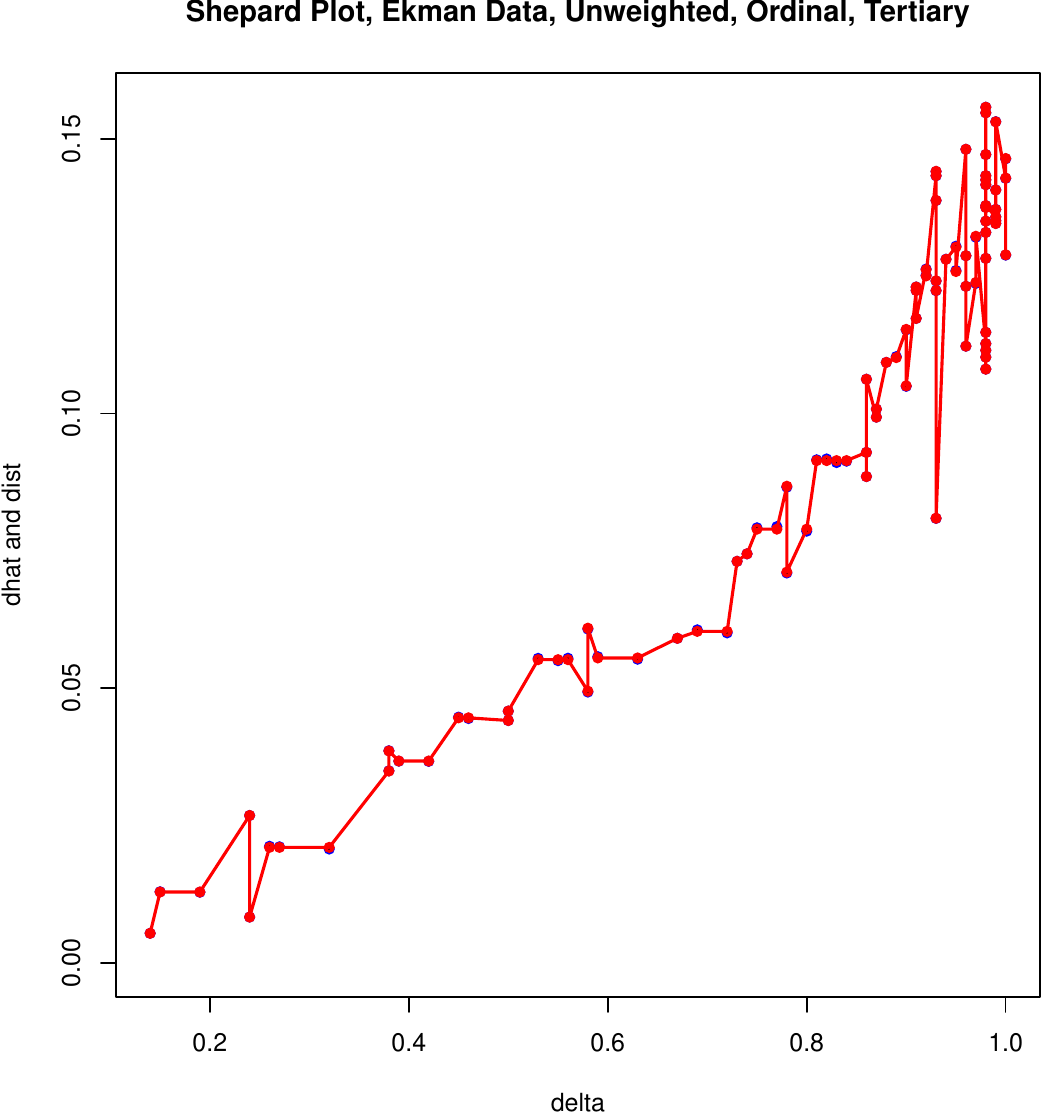}}
\end{center}

\begin{center}
\pandocbounded{\includegraphics[keepaspectratio]{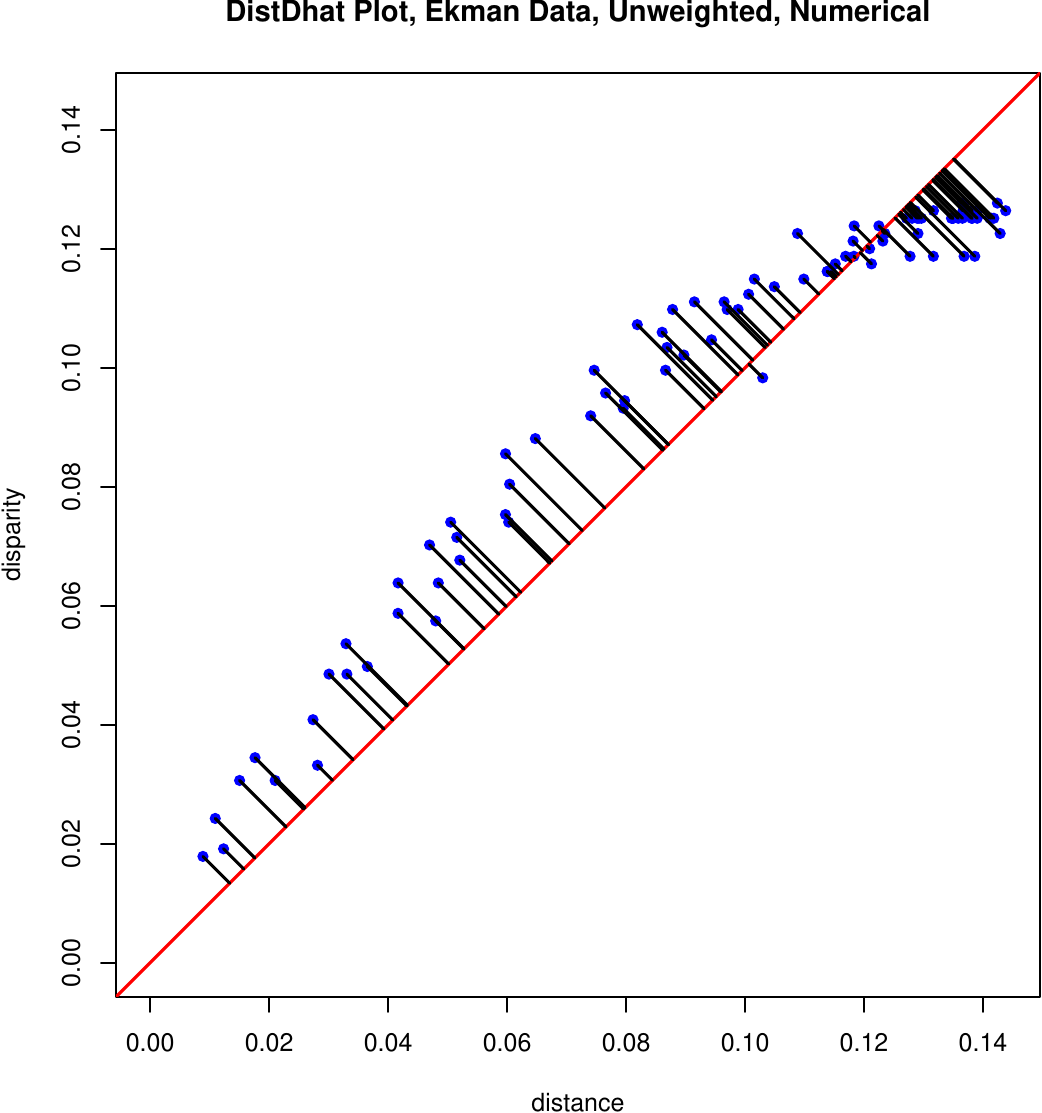}}
\end{center}

\begin{center}
\pandocbounded{\includegraphics[keepaspectratio]{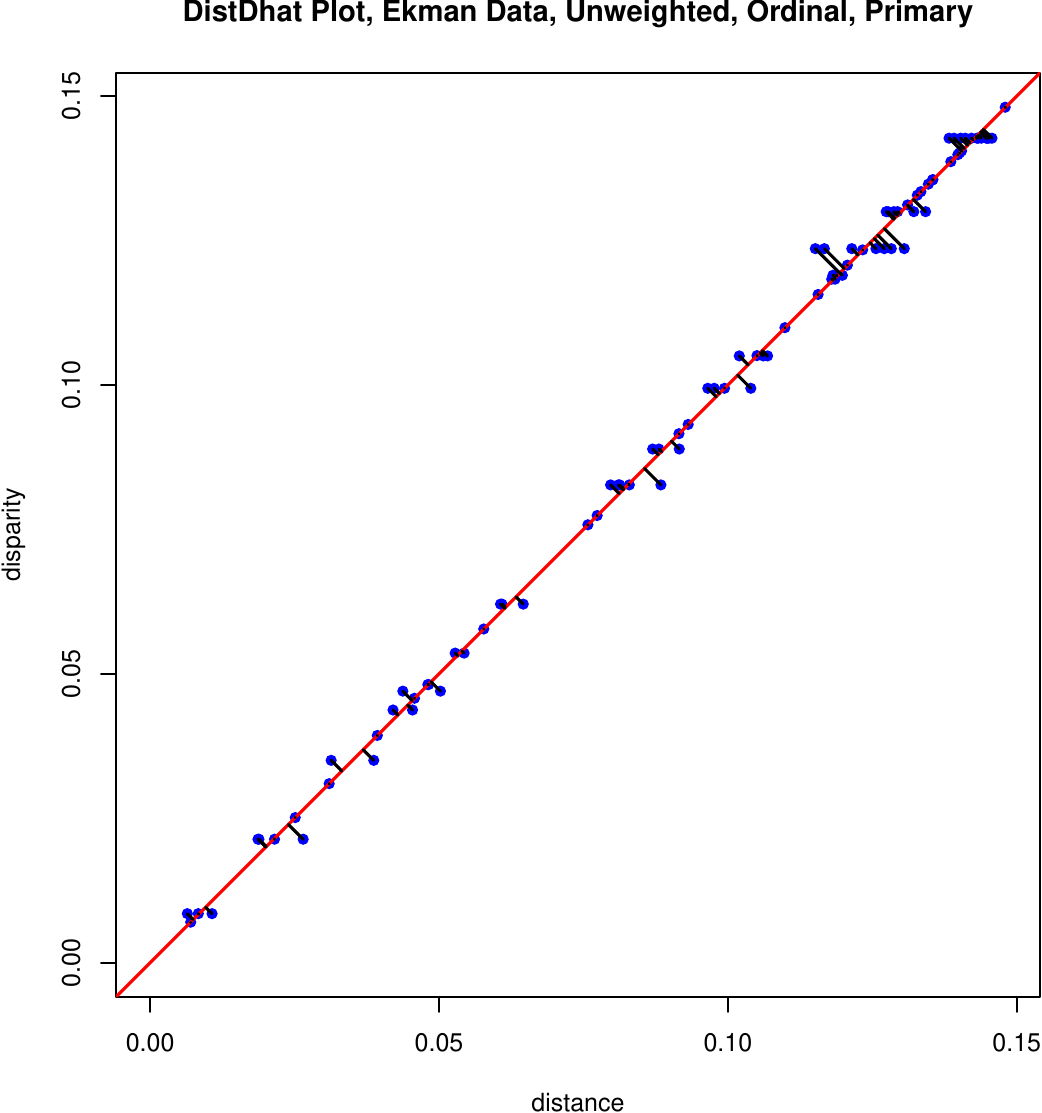}}
\end{center}

\begin{center}
\pandocbounded{\includegraphics[keepaspectratio]{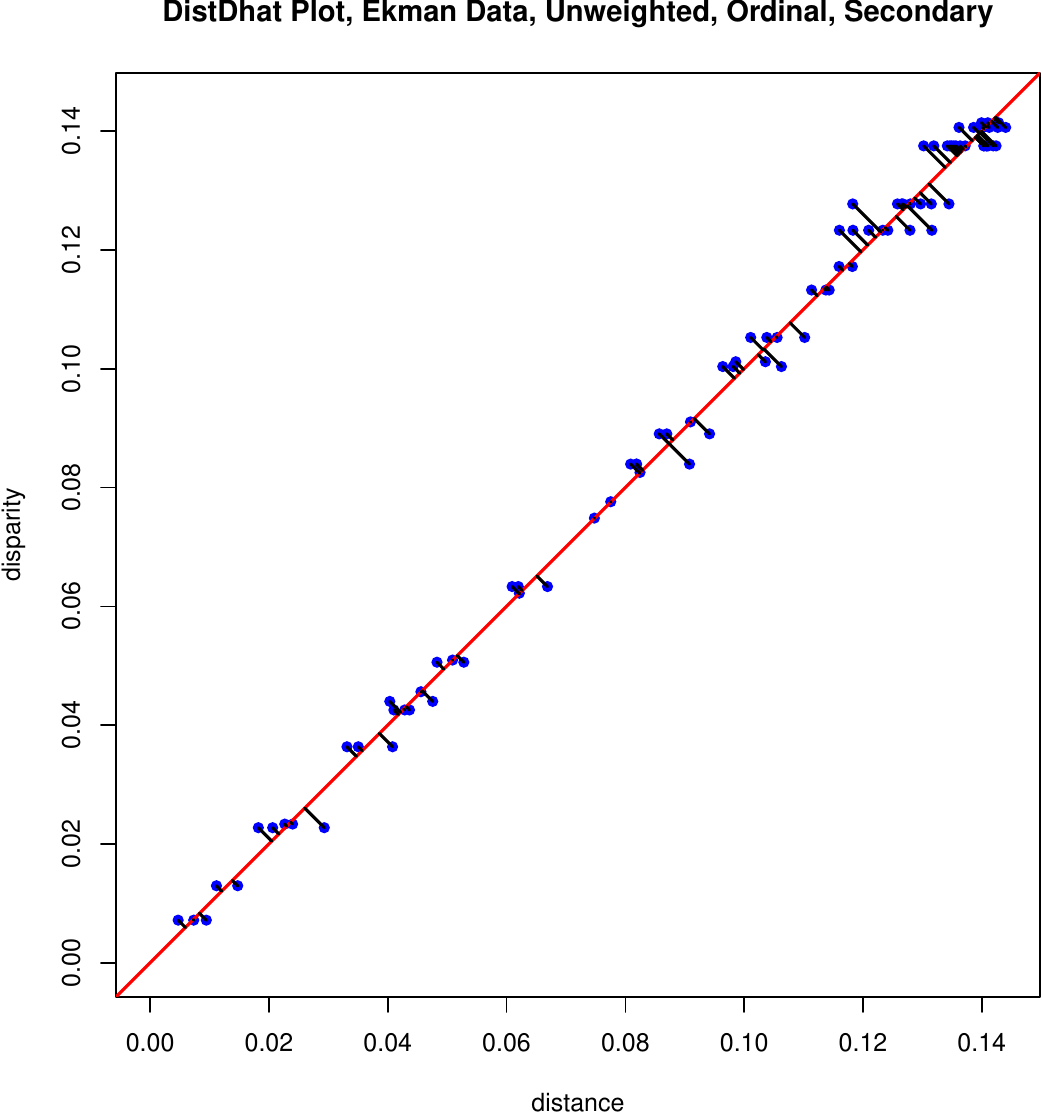}}
\end{center}

\begin{center}
\pandocbounded{\includegraphics[keepaspectratio]{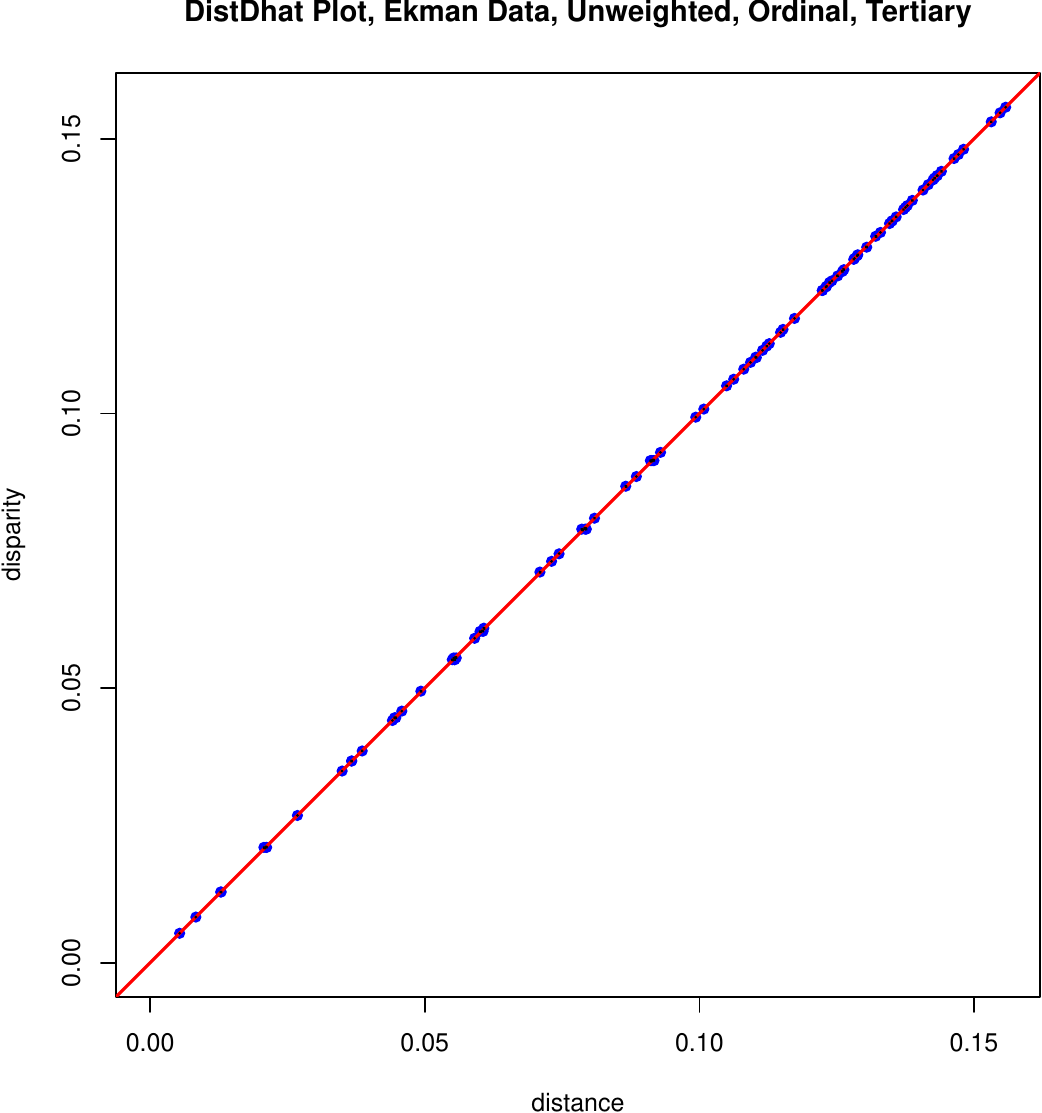}}
\end{center}

\begin{center}
\pandocbounded{\includegraphics[keepaspectratio]{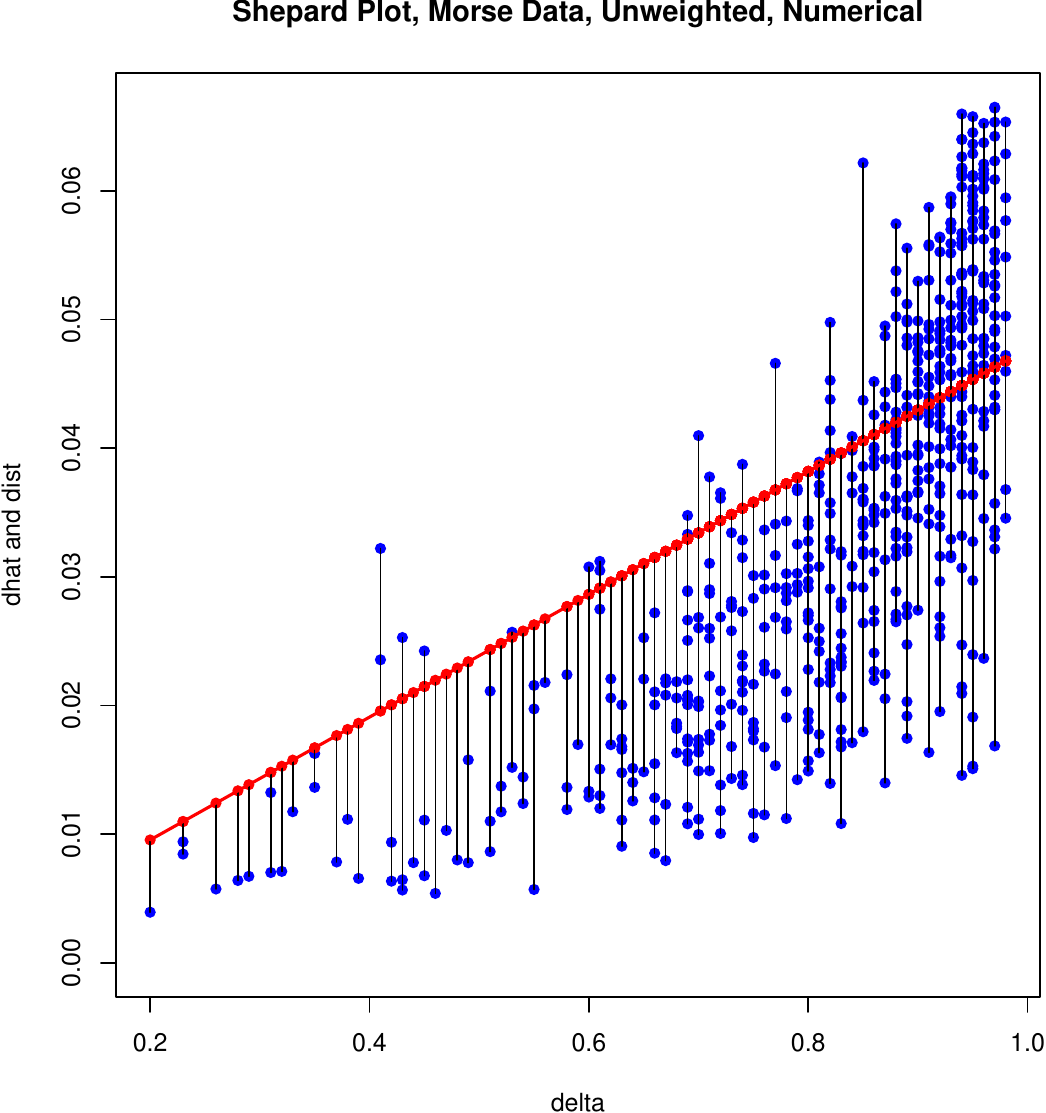}}
\end{center}

\begin{center}
\pandocbounded{\includegraphics[keepaspectratio]{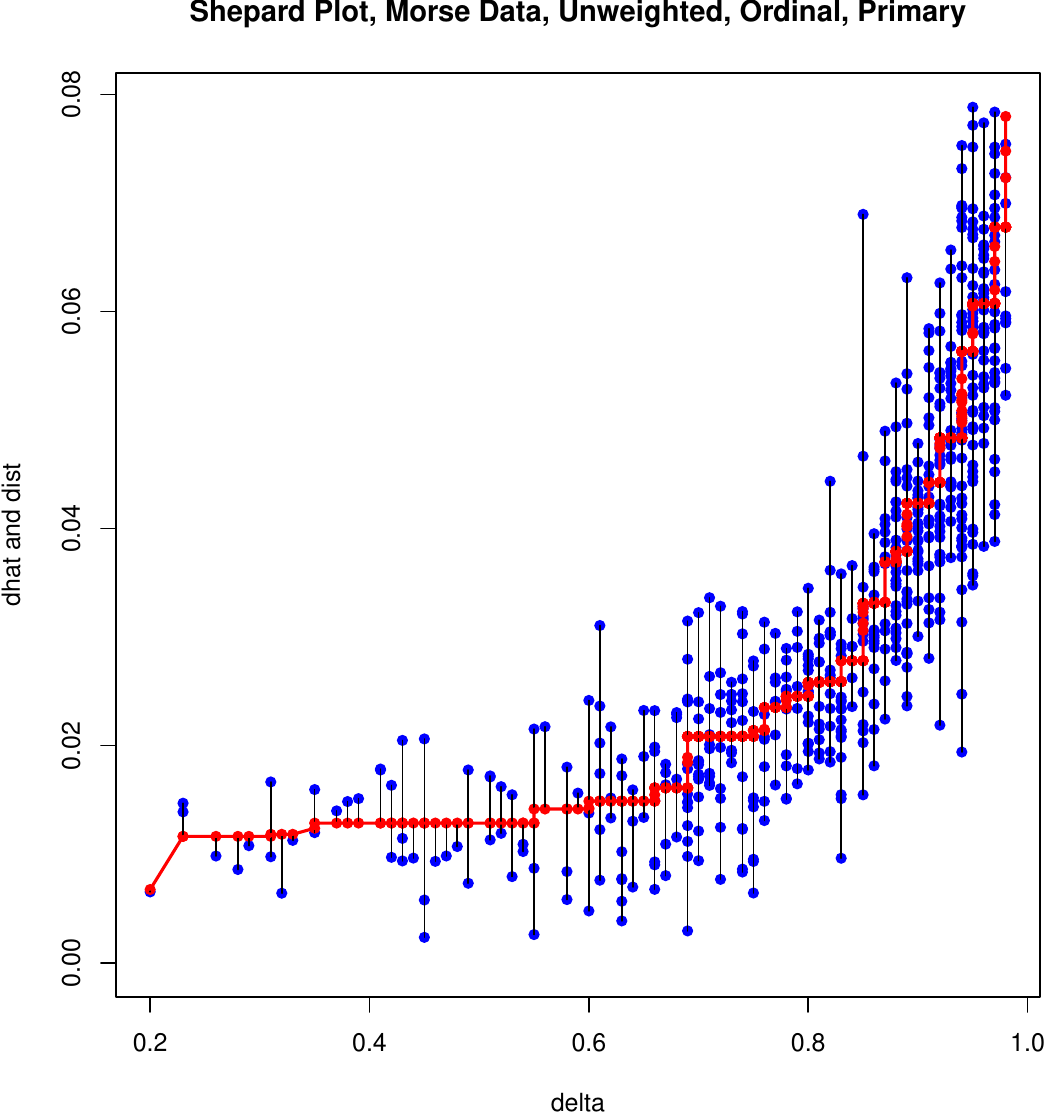}}
\end{center}

\begin{center}
\pandocbounded{\includegraphics[keepaspectratio]{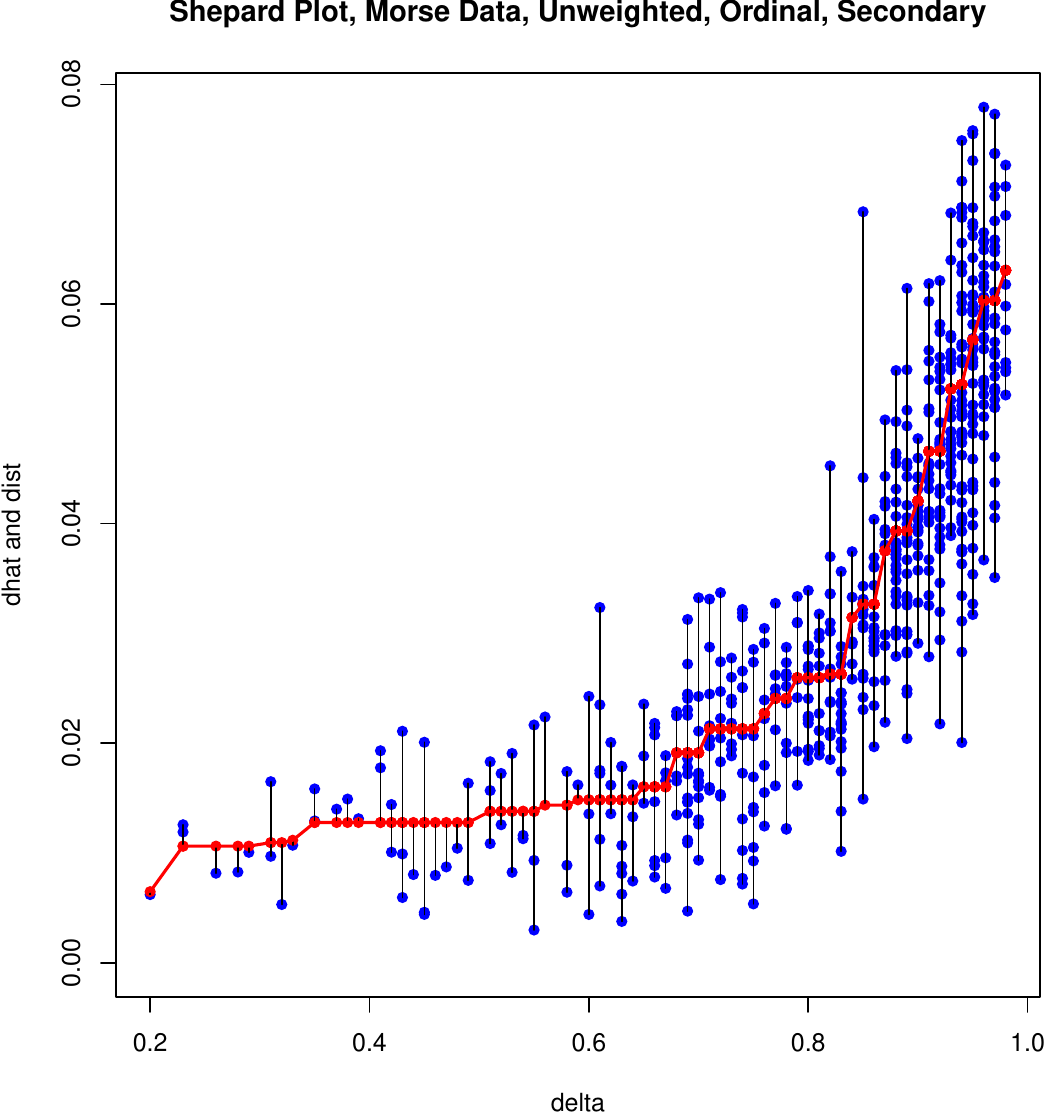}}
\end{center}

\begin{center}
\pandocbounded{\includegraphics[keepaspectratio]{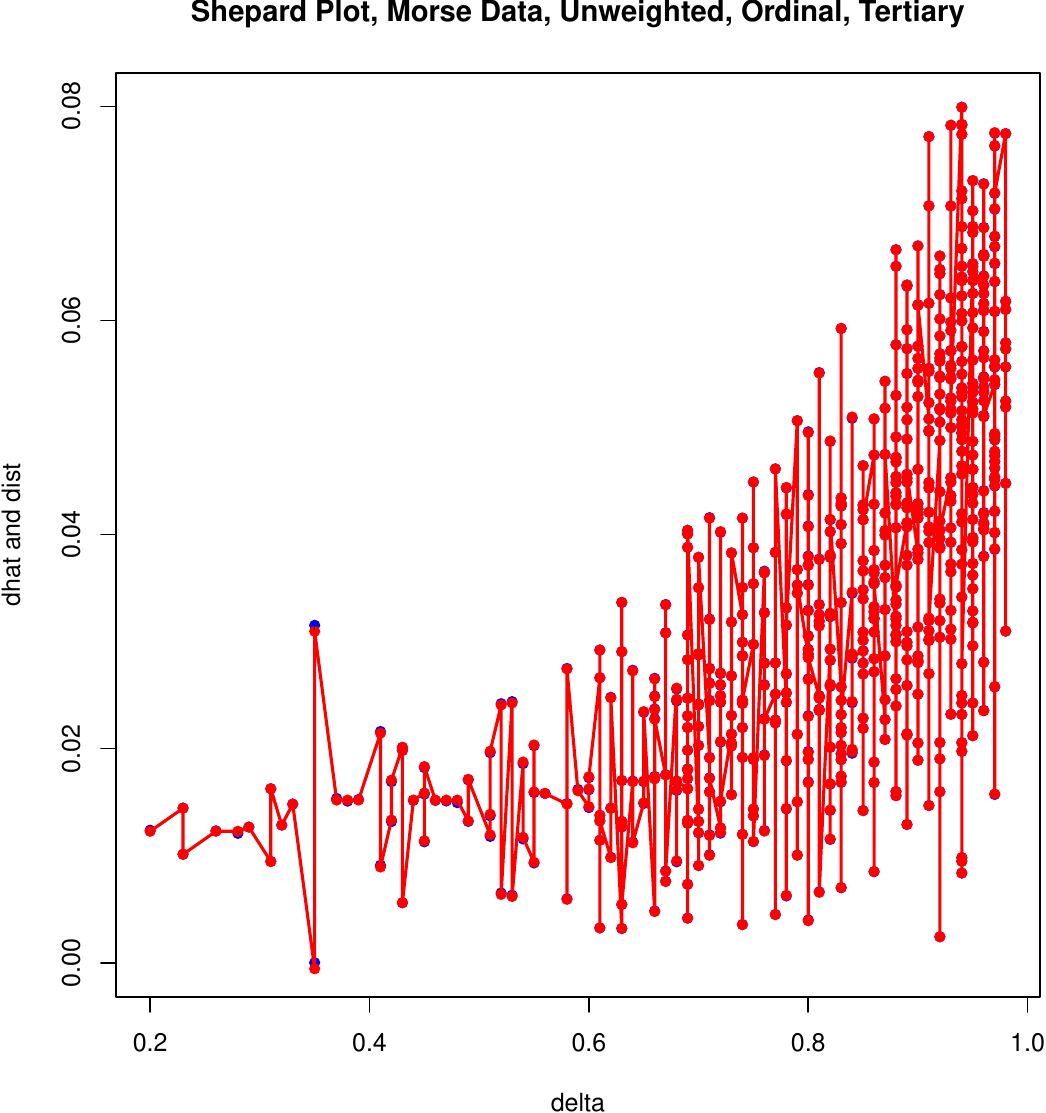}}
\end{center}

\begin{center}
\pandocbounded{\includegraphics[keepaspectratio]{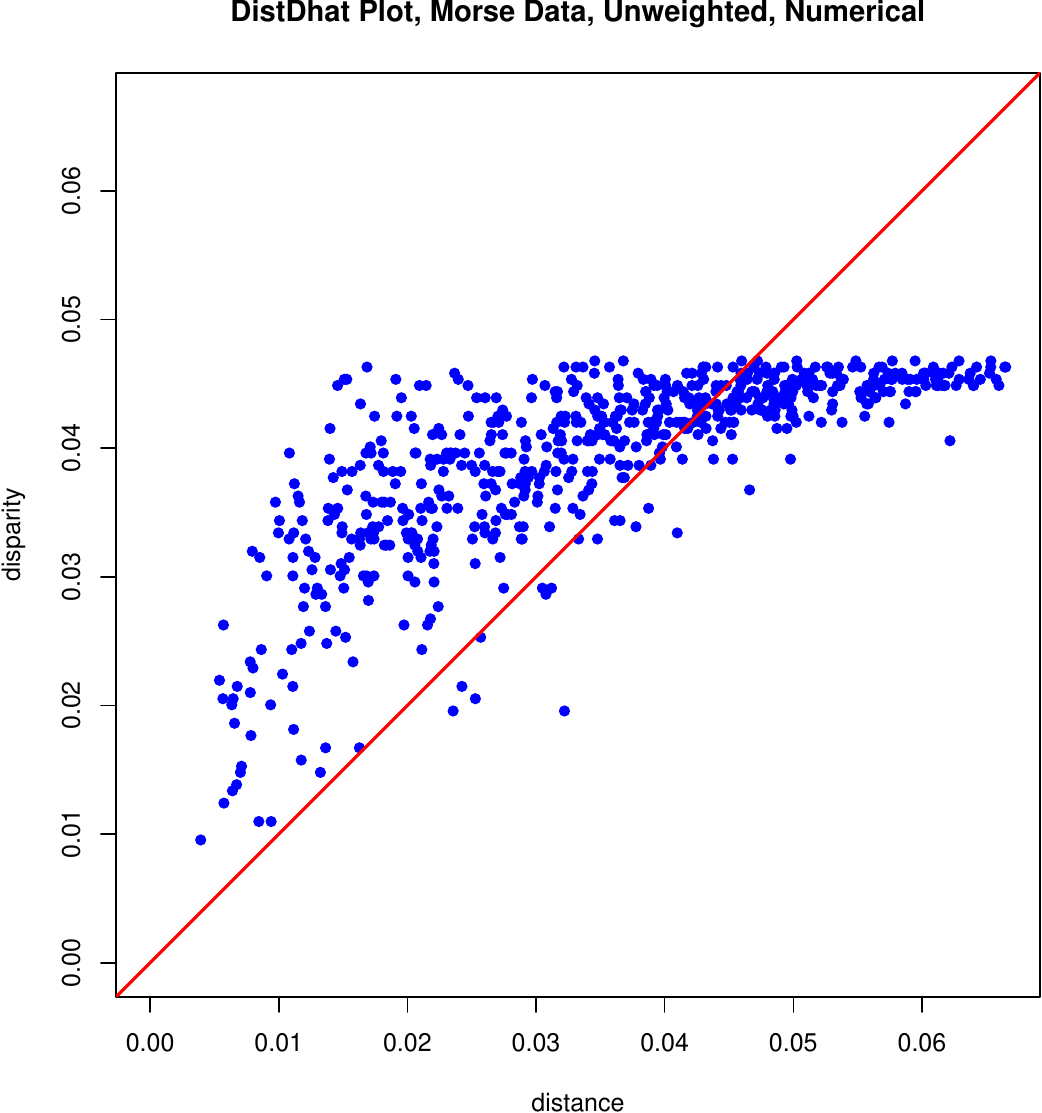}}
\end{center}

\begin{center}
\pandocbounded{\includegraphics[keepaspectratio]{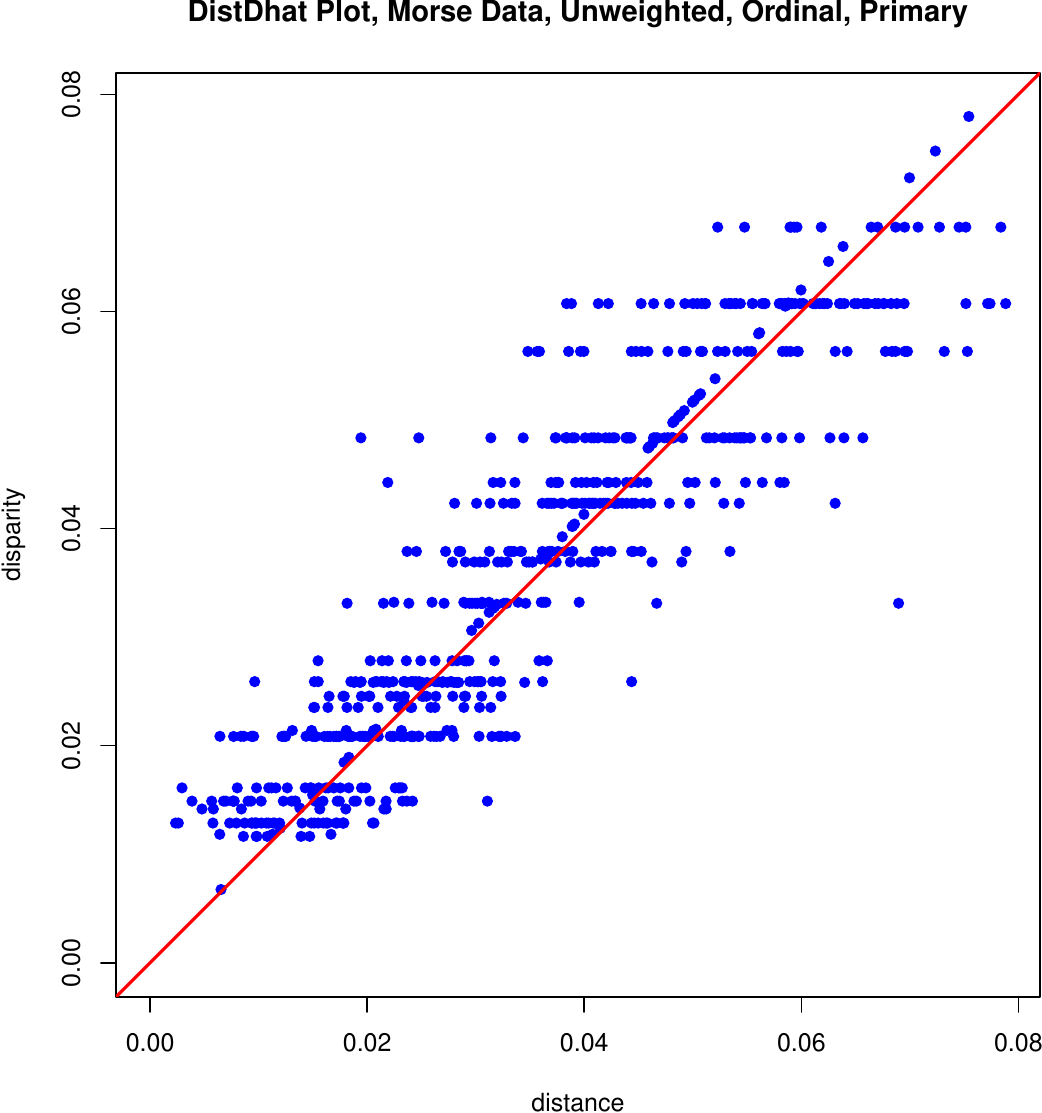}}
\end{center}

\begin{center}
\pandocbounded{\includegraphics[keepaspectratio]{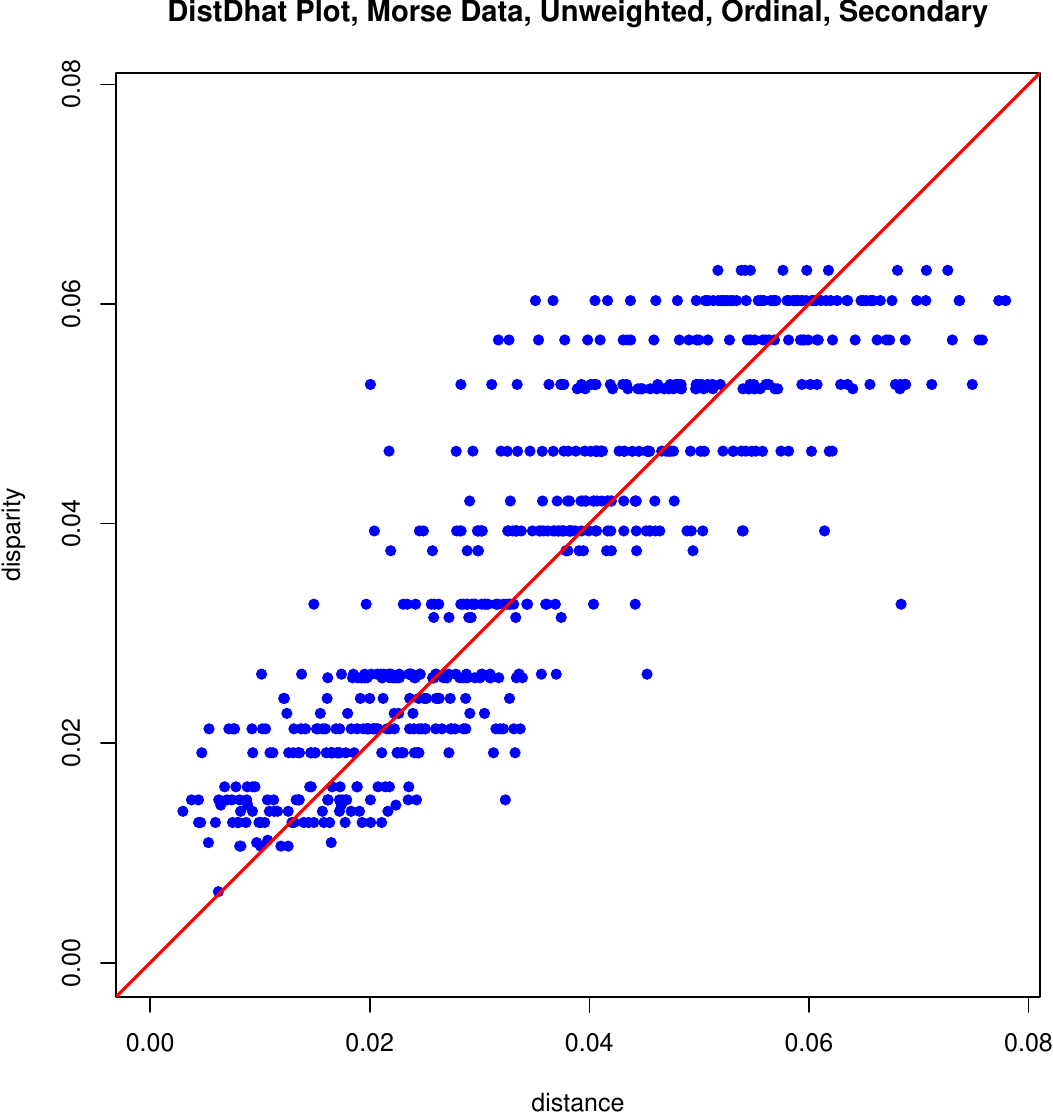}}
\end{center}

\begin{center}
\pandocbounded{\includegraphics[keepaspectratio]{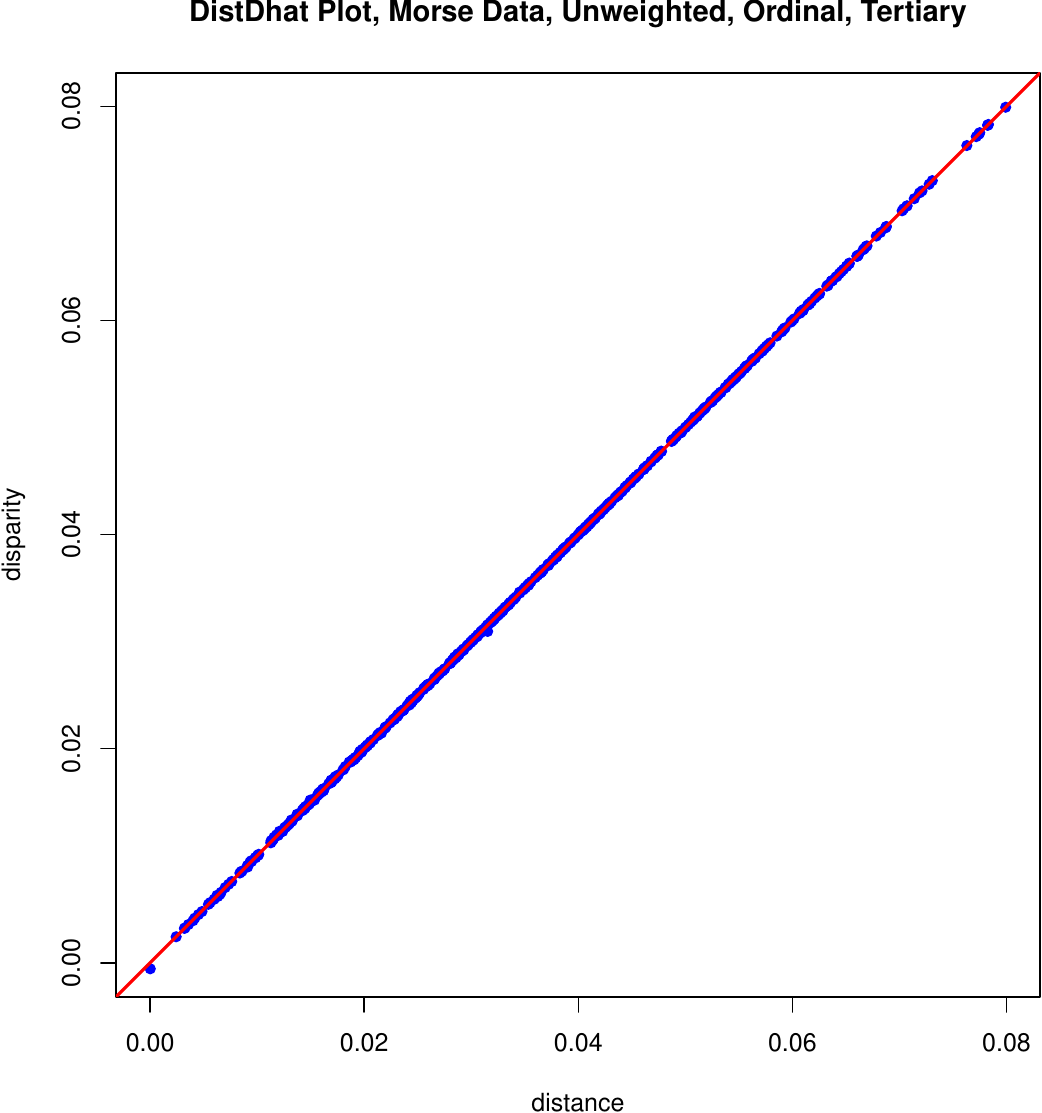}}
\end{center}

\section*{References}\label{references}
\addcontentsline{toc}{section}{References}

\phantomsection\label{refs}

\end{document}